\documentclass[twocolumn]{aastex631}

\usepackage{epsfig}
\usepackage{url}
\usepackage{hyperref}

\usepackage[normalem]{ulem} 

\usepackage{latexsym}
\usepackage{epsfig}
\usepackage{amsmath}
\usepackage{amssymb}
\usepackage{wasysym}
\usepackage{graphicx}
\usepackage{dcolumn}
\usepackage{verbatim}
\usepackage{enumerate,mdwlist}

\usepackage[titletoc]{appendix}

\usepackage{amsfonts}
\usepackage{fancyvrb} 
\usepackage{tikz} 
\usetikzlibrary{calc}
\makeatletter\let\splitbox\@undefined\makeatother 
\usepackage[export]{adjustbox}

\usepackage[normalem]{ulem}

\usepackage[inline, final]{showlabels}
\usepackage{tensor}

\usepackage{listings}
\usepackage{xcolor}
\IfFileExists{fontawesome5.sty}{\usepackage{fontawesome5}}{} 

\definecolor{comment_red}{rgb}{0.5, 0, 0}
\newcommand{\ee}{\text{e}} 
\newcommand{\ii}{\text{i}}
\newcommand{\di}{\text{d}}

\makeatletter
\g@addto@macro\bfseries{\boldmath}  
\makeatother

\renewcommand{\v}[1]{\boldsymbol{#1}}

\newcommand{\ba}{\begin{eqnarray}}
\newcommand{\ea}{\end{eqnarray}}

\newcommand{\glow}{\texttt{GLoW}}

\newcommand{\ymacro}{y_{\rm macro}}
\newcommand{\Msrctot}{M_{\rm tot}^{\rm src}}

\definecolor{grey}{rgb}{0.4,0.4,0.4}
\definecolor{dullmagenta}{rgb}{0.4,0,0.4}
\definecolor{darkblue}{rgb}{0,0,0.4}
\definecolor{midblue}{rgb}{0,0,0.5}
\definecolor{midred}{rgb}{0.5,0,0}
\definecolor{orange}{rgb}{1,0.5,0}
\definecolor{lightbrown}{rgb}{0.75,0.5,0.25}
\definecolor{tan}{cmyk}{0.14,0.42,0.56,0}
\definecolor{djunglegreen}{cmyk}{0.99,0,0.52,0}
\definecolor{lightgreen}{rgb}{0,1,0}
\definecolor{olivegreen}{cmyk}{0.64,0,0.95,0.40}
\definecolor{midgreen}{rgb}{0.0,0.675,0.0}
\definecolor{darkgreen}{rgb}{0,0.5,0}
\definecolor{ceruleanblue}{rgb}{0.0, 0.2, 0.7}
\definecolor{burgundy}{rgb}{0.5, 0.0, 0.13}
\definecolor{hvred}{RGB}{186,12,47}

\hypersetup{
    colorlinks=true,
    linkcolor=ceruleanblue,
    filecolor=ceruleanblue,      
    urlcolor=midblue,
    citecolor=burgundy,
}

\usepackage[all]{xy} 
\usepackage{amsfonts}

\makeatletter
\def\l@subsubsection#1#2{}
\makeatother

\begin{document}

\title{Across the Universe: \\ GW231123 as a Magnified and Diffracted Black Hole Merger}

\author{Srashti Goyal}
\email{srashti.goyal@aei.mpg.de}
\affiliation{Max Planck Institute for Gravitational Physics (Albert Einstein Institute) \\
Am Mühlenberg 1, D-14476 Potsdam-Golm, Germany}

\author{Hector Villarrubia-Rojo}
\email{hectorvi@ucm.es}
\noaffiliation{}

\author{Miguel Zumalac\'arregui}
\email{miguel.zumalacarregui@aei.mpg.de}
\affiliation{Max Planck Institute for Gravitational Physics (Albert Einstein Institute) \\
Am Mühlenberg 1, D-14476 Potsdam-Golm, Germany}

\begin{abstract}
GW231123 is the most massive binary black hole merger observed to date by the LIGO–Virgo–KAGRA detectors, with a reported total mass of $190\text{--}265\,M_\odot$ (90\% c.l.). Such a high observed mass can result from the combined effects of cosmological redshift and gravitational magnification if the source is gravitationally lensed. In addition, compact objects such as stellar remnants can induce wave-optics diffraction, imprinting characteristic distortions on the signal.

We present a multi-scale lensing analysis of GW231123 that incorporates diffraction by a compact microlens embedded in an external gravitational potential. The data favor a lensed interpretation with a conservative false-alarm probability of $\lesssim 1\%$. The inferred total source mass shifts to $100\text{--}180\,M_\odot$, reducing the need for extreme spins and tensions between waveform models. We reconstruct both source and lens parameters, including the microlens mass ($190\text{--}850\,M_\odot$), its projected offset, and the amplitude and orientation of the external potential. 

Assuming a galaxy-scale macrolens modeled as a singular isothermal sphere, we infer external magnification that places the source at redshift $z\sim0.7\text{--}2$ and predict a $\sim55\%$ probability of forming an additional detectable image. This analysis demonstrates how current gravitational-wave observations probe the wave-optics lensing regime, enabling joint inference of compact lenses and macroscopic gravitational potentials, including bounds on compact dark matter. Continued searches for additional macroimages and diffraction by stellar fields will further test GW231123 as a lensed event candidate.
\end{abstract}

\keywords{Gravitational lensing (670) --- Gravitational microlensing (672) --- Strong gravitational lensing (1643) --- Gravitational waves (678) --- Dark matter (353) --- Primordial black holes (1292) --- Astrophysical black holes (98) --- Stellar mass black holes (1611) --- Intermediate-mass black holes (816)}

\date{\today}


\section{Introduction}
On November 2023 the LIGO detectors recorded gravitational waves (GWs) from the coalescence of two black holes with total mass of 190\text{--}265 $M_\odot$ \citep{LIGOScientific:2025rsn, advligo}. This event, known as GW231123\_135430 (hereafter referred as GW231123), represents the most massive binary detected so far, consistently heavier than previously found signals~\citep{KAGRA:2021vkt,gwtc4catalog}. 
The standard analysis indicates a preference for unusually high component spins, beyond the values expected even for second-generation merger remnants~\citep{Fishbach:2017dwv,Borchers:2025sid}. 
Analyses using different waveform approximants produce mutually inconsistent results, pointing towards waveform systematics or missing physical effects. 
These unique properties have led to many investigations into the origin of GW231123~\citep{Stegmann:2025cja,Tanikawa:2025fxw,Liu:2025ogx,Bartos:2025pkv,Delfavero:2025lup,Li:2025pyo,Li:2025fnf,Paiella:2025qld,Wang:2026yjk,Croon:2025gol,DeLuca:2025fln,Cuceu_2026}. GW231123 was also reported as showing the ``strongest
support for distorted lensed signals seen so far''~\citep{LIGOScientific:2025rsn}, a hypothesis that we consider in detail here.

An unusually massive binary coalescence can be explained by the combination of two effects: the cosmological redshift reduces the observed frequency, making the binary appear heavier $m^{\rm det}=(1+z)m^{\rm src}$, while gravitational magnification increases the signal's amplitude, making the source appear closer $d_L^{\rm det}=d_L/\sqrt{\mu}$, where $z$ and $\mu$ are the source's redshift and magnification ~\citep{Dai:2016igl,Oguri:2018muv,Oguri:2019fix,Broadhurst:2018saj,Diego:2021fyd,Pang:2020qow,Smith:2022vbp,Magare:2023hgs,Farah:2025ews}. 
Magnified GWs travel through the stellar fields of galaxies and clusters, known as microlenses, which may distort the signal in observable ways~\citep{Christian:2018vsi,Dai:2018enj,Diego:2019lcd,Diego:2019rzc,Cheung:2020okf,Lewis:2020asm,Yeung:2021chy,Mishra:2021xzz,Oguri:2022zpn,mishrapopbiases2023,Meena:2023qdq,Palencia:2023kne,Shan:2024min,Chakraborty:2024mbr,Seo:2025dto,Smith:2025axx,Su:2025xry}. 
In addition to expected stellar objects, GWs will be also sensitive to the small-scale distribution of dark matter~\citep{Jung:2017flg,Choi:2021bkx,Oguri:2020ldf,Urrutia:2021qak,Fairbairn:2022xln,Tambalo:2022wlm,Savastano:2023spl,GilChoi:2023qrz,Urrutia:2024pos,Zumalacarregui:2024ocb,Brando:2024inp,Vujeva:2025nwg,Choi:2026lsa}.

GWs undergo lensing diffraction in the wave-optics regime, as the gravitational radius of typical stellar objects is comparable to the signal's wavelength~\citep{Takahashi:2003ix,Leung:2023lmq}. Advances in wave-optical lensing computations~\citep{Feldbrugge:2019fjs,Tambalo:2022plm,Shan:2022xfx,Yeung:2024pir,Villarrubia-Rojo:2024xcj} and parameter estimation~\citep{Wright:2021cbn,Cheung:2024ugg,Our_methods_paper} now enable the search for these signatures, although inference has only been possible assuming a symmetric potential, e.g.~an isolated point lens. 
Unambiguous evidence for lensed GWs has not yet been reported~\citep{Hannuksela:2019kle,Dai:2020tpj,LIGOScientific:2021izm,LIGOScientific:2023bwz,Chakraborty:2025maj}, but many studies indicate that lensed GWs are expected in the upcoming upgrade of the LIGO-Virgo-KAGRA network~\citep{Ng:2017yiu,Li:2018prc,Smith:2022vbp,Wierda:2021upe,network}. 

The inevitability of gravitational lensing and unusual properties of GW231123 warrant a detailed examination of the lensing hypothesis. Ref.~\citep{lvklensing} found GW231123 as an outlier in GWTC-4 during their searches for microlensing, assuming a point lens mass model (see also Refs.~\cite{Shan:2025dcd,Chakraborty:2025pxt,Chan:2025kyu,Hu:2025lhv,Cheung:2026pky}).
In this Letter, we report on the first analysis using a non-symmetric lens model, which describes diffraction by a point lens embedded in an anisotropic gravitational potential, as expected in a strongly lensed system.
We show how distortions of the waveform can probe vastly different scales in the system, from stellar to galactic.
While GW231123 provides a particularly compelling case study, the multi-scale framework developed here is broadly applicable and will be relevant as additional data and analyses further test the lensing interpretation of this and other GW events.

We will first introduce the embedded point lens model and its connection to the macroscopic lens. Then we present Bayesian inference results followed by a discussion of the source and microlens properties and strong-lensing predictions. We conclude by discussing the limitations of our analysis and possible directions for future research.

\section{Diffraction by a point mass embedded in an external potential}\label{sec:embedded_pl}
\begin{figure}
    \centering
    \includegraphics[width=0.99\linewidth]{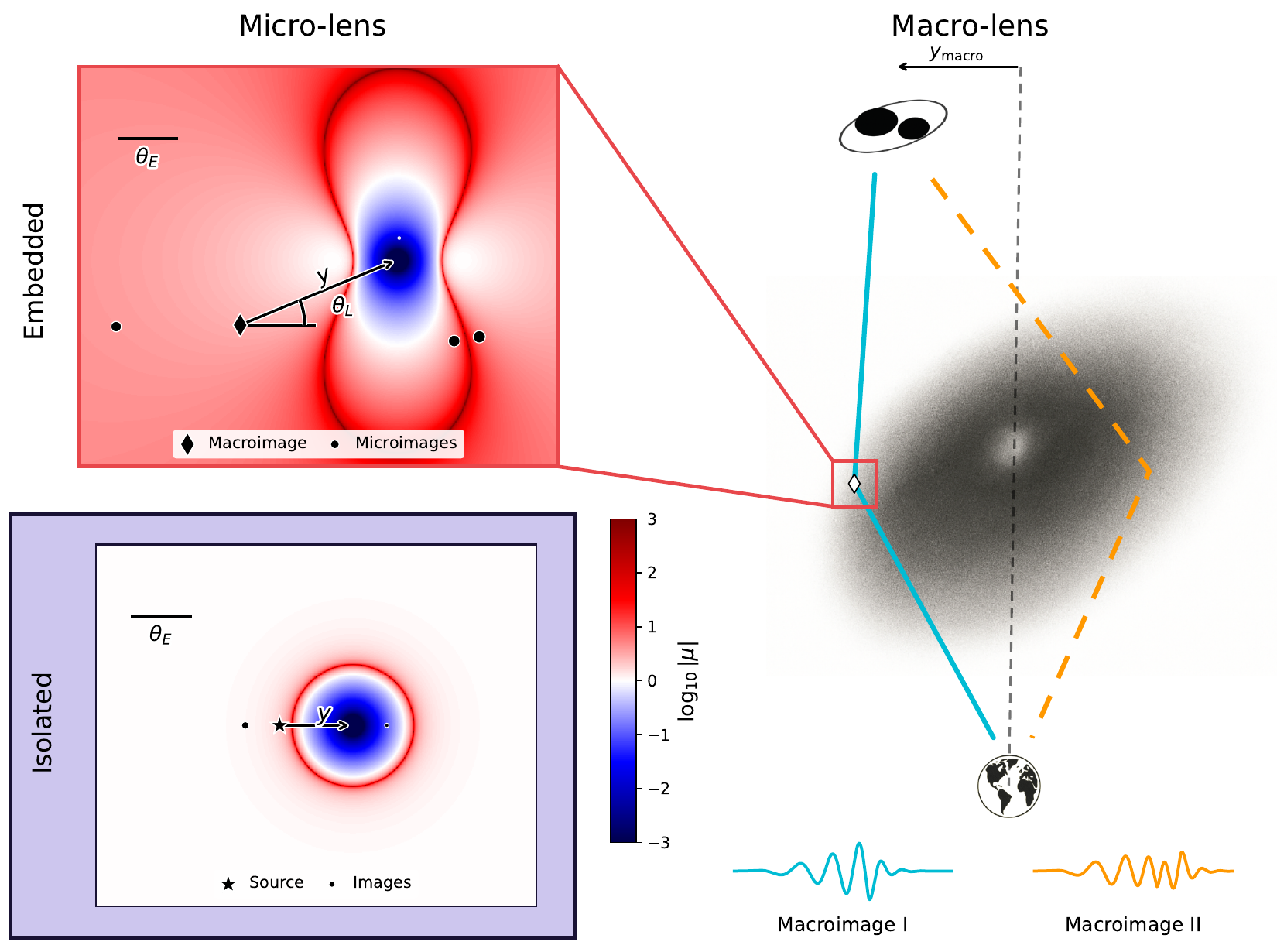}
    \caption{Diffracted and magnified GWs. Top left: A small-scale point-lens in an external potential diffracts the signal. The external potential distort and enlarge the Einstein ring and produces additional microimages (color shows their magnification). 
    This represents the local region near a macroimage produced by a massive object, e.g.~ a galaxy, shown in the right panel. The offset between the lens and source ($\ymacro$) determines whether additional images (dashed line) form. In contrast, the isolated point lens (bottom left panel) neglects effects from the macroscopic potential and only produces two microimages.
    }
    \label{fig:lensing_diagram}
\end{figure}

We will consider a symmetric lens (hereafter microlens) embedded in an external potential (hereafter macro-potential~\cite{Chang:1979zz,An:2006bq}). The system is characterized by the Fermat potential, which is the sum of time delays due to the geometric deflection, the external macro-potential and the microlens: $\phi(\v{x}) = \frac{1}{2}|\v{x}|^2 - \psi_{\rm ext}(\v{x})- \psi_m(\v{x})$. Specializing to a quadratic macro-potential and a point lens yields,
  \begin{align}
    \phi(\v{x}) &=
    \frac{1}{2}\left[(1-\kappa -\gamma)x_1^2+(1-\kappa +\gamma)x_2^2\right] 
                 \nonumber  \\
                   &\quad- \frac{1}{2}\log\left[(x_1-y\cos\theta_L)^2+ (x_2-y\sin\theta_L)^2\right]
                   \,.
\end{align}
The macro-potential is characterized by convergence $\kappa$ (isotropic) and shear $\gamma\geq 0$ (directional), that produce
a macroimage with magnification $\mu_{\rm macro} = \left((1-\kappa)^2-\gamma^2\right)^{-1}$. We choose coordinates
centered at the apparent source position (macroimage), aligned with the external shear. 
The point mass microlens is located at a distance $y$ from the apparent source position and at an angle $\theta_L \in [0,\pi/2)$ relative to the external shear.
Finally, we have rescaled the coordinates by the Einstein radius of the microlens $\xi_0 = \sqrt{4GM_{Lz} d_{\rm eff}}\approx 0.14\,{\rm pc}\sqrt{\frac{M_{Lz}}{100M_\odot}\frac{d_{\rm eff}}{1{\rm Gpc}}}$, where $M_L$ is the microlens mass, $M_{Lz}\equiv M_L(1+z_L)$ its redshifted value, and $d_{\rm eff}\equiv D_L D_{LS}/[(1+z_L)D_S]$ an effective lensing distance, with $D_L$, $D_S$, $D_{LS}$ the angular diameter distances to the lens, to the source and between the lens and the source. For details on lensing conventions see \cite{Schneider1999-tk,Villarrubia-Rojo:2024xcj}.
Fig.~\ref{fig:lensing_diagram} shows the geometry of the microlens embedded in an external potential, including the extended lens around the system. The simpler isolated point lens ($\kappa=\gamma=0$) is shown for comparison.

Lensing diffraction is characterized by the \textit{amplification factor}, the ratio of the lensed to unlensed frequency-domain waveforms $F(f)\equiv \frac{\tilde h(f)}{\tilde h_0(f)}$. It is given by
\begin{equation}\label{eq:Fw_def}
   F(w) = \frac{w}{2\pi \ii}\int\di^2 x\exp\Big(\ii w\left({\phi}(\v{x})-\phi_0\right)\Big)\,,
\end{equation}
where $\phi_0$ is the minimum of $\phi(\v{x})$ and the GW signal frequency $f$ has been rescaled by the \textit{redshifted lens mass} $M_{Lz}$, resulting in a dimensionless frequency,
\begin{equation}
w \equiv 8\pi GM_{Lz}f \sim 1.2 \left(\frac{M_{Lz}}{100 M_\odot}\right)\left(\frac{f}{100{\rm Hz}}\right)\,.
\end{equation}
Point masses satisfy $M_{Lz} = (1+z_L)M_L$, where $M_L$ is the mass of the lens and $z_L$ is its redshift, but extended lenses have different relations to the total mass~\citep{Tambalo:2022wlm}. 

For any given lens, the amplification factor can be split as
\begin{equation}
    F(w) = \sum_I \sqrt{\mu_I}\,\ee^{\ii w \phi_I + \ii n_I} + F_{\rm WO}(w)\,.
\end{equation}
The sum runs over \textit{geometric optics} images $I$ (stationary points of the Fermat potential $\v{\nabla}\phi(\v{x}_I)=0$), each characterized by a magnification $\mu_I$, time delay $\phi_I \equiv \phi(\v{x}_I)$ and phase $n_I = 0,\pi/2, \pi$ if the image is a local minimum, saddle point or maximum of $\phi(\v{x})$~\citep{Dai:2017huk,Ezquiaga:2020gdt}. Fig. \ref{fig:lensing_diagram} shows an example of micro-images and magnification for an isolated and embedded point-lens. $F_{\rm WO}(w)$ encodes diffractive wave-optics effects that vanish when $w\to \infty$.
For details on wave-optics lensing and methods to evaluate $F(w)$ see~\cite{Villarrubia-Rojo:2024xcj}. 

We choose an external potential that forms a type I macroimage, i.e. a minimum of the Fermat potential, corresponding to $\kappa+\gamma<1$ and $\kappa < 1$. The macro-potential can be simplified using the mass sheet transformation to remove the convergence $ \kappa\to \tilde \kappa = 0$~\citep{mass_sheet_degeneracy_original,1988ApJ_Gorenstein_MSD,Wagner:2018rae,Wagner:2019azs}.
This rescales $\gamma\to\tilde{\gamma}\equiv \gamma/(1-\kappa)$, and $y\to \tilde{y}\equiv \sqrt{1-\kappa}\;y$. The transformed amplification factor \eqref{eq:Fw_def} is
\begin{align}
    F(w;\kappa, \gamma, y,\theta_L) = \frac{1}{1-\kappa}F(w;0,\tilde{\gamma},\tilde{y},\theta_L)\,.
    \label{eq:F_mass_sheet}
\end{align} 
The rescaling of $F(w)$ implies the following relation between the real ($d_L$) and inferred ($\tilde{d}_L$) luminosity distance: $ d_L \to \tilde d_L \equiv (1-\kappa) d_L$. 
Consequently, the inferred macro-magnification is $\tilde \mu_{\rm macro} = \left(1-\tilde\gamma^2\right)^{-1}=\mu_{\rm macro}\left(1-\kappa\right)^{2}$.
Therefore, determining the true source distance and macro-magnification requires more information on the macrolens, e.g.~a relation between $\kappa$ and $\gamma$ that breaks the mass-sheet degeneracy, Eq.~\eqref{eq:F_mass_sheet}.

We will discuss general results ($\kappa$ unspecified) and also assume that the macroscopic lens is a single galaxy ($\kappa$ fixed by $\tilde \gamma$) modelled as a \textit{singular isothermal sphere} (SIS) with density $\propto 1/r^2$. Under this assumption, the external potential satisfies
\begin{equation}\label{eq:macro_SIS}
    \kappa = \gamma = \frac{\tilde \gamma}{1+\tilde \gamma}  \quad \text{(SIS)}  \,,
\end{equation}
at the macroimage positions (the elliptical version of the SIS preserves this relation~\cite{Kormann:1994}). 
Isothermal profiles are a very good approximation to the central region of galactic-scale lenses~\citep{Oguri:2001qu}, a plausible macrolens for GWs at current sensitivity~\citep{Hilbert:2007ny,Diego:2018fzr,Robertson:2020mfh}.
The SIS relation Eq.~\eqref{eq:macro_SIS} will be used show how GWs can inform our understanding of a macrolens: future analyses can be generalized to a more general relationship between $\kappa$ and $\tilde\gamma$ due to ellipticity, external shear, line-of sight effects~\citep{Keeton:1996tq,CaganSengul:2020nat,Fleury:2021tke}.

SIS shear and convergence are determined by the macroscopic offset $\ymacro$ (cf.~Fig.~\ref{fig:lensing_diagram}) as $\kappa = \gamma = \frac{1}{2(\ymacro\pm 1)}$, where $+,-$ refer to the first and second image.
We will assume that GW231123 corresponds to the first macroimage (+, type I) at a local minimum of the macro-potential, where $\tilde\gamma \in [0,1)$. A local minimum is always produced for any source-lens-observer alignment and enhances the signal's amplitude~\citep{Blandford:1986zz}.
Under the SIS hypothesis, an additional image forms at a saddle point of the macro-potential ($-$, type II) when $\ymacro<1$.
Although model-dependent, the SIS macrolens assumption allows us to infer the true luminosity distance to the source and microlens impact parameter, as well as predict the existence and properties of additional images.

The analysis below demonstrates the capabilities of this framework when applied to real GW data, using GW231123 as a particularly informative example. The microlens model represents the most complete wave-optics treatment robust enough for full Bayesian parameter estimation of gravitational-wave signals. It represents a first approximation to a complex stellar field, capturing the regime in which a dominant object is embedded in a larger-scale gravitational potential.

The macroscopic lens is modelled with a deliberately simple, analytically tractable prescription. The GW data constrain only the \emph{local} quantities ---the reduced shear and the embedded microlens parameters--- which set the chromatic diffraction pattern and are invariant under the mass-sheet transformation. They do not fix the absolute macro-magnification, which is a model-dependent conversion. The SIS is used solely as a closure that fixes the convergence at the image position ($\kappa=\gamma$), turning the locally inferred reduced shear into illustrative estimates of the source redshift, the physical microlens offset, and the number and brightness of additional macroimages.

A more realistic macro-model ---an elliptical isothermal lens, external shear, or a distribution of $\kappa$ at fixed reduced shear--- would leave the local microlensing inference unchanged, but can produce a substantially larger absolute magnification, especially near fold or cusp caustics~\citep{Miralda-Escude:1991,Diego:2018fzr}. Because the mass-sheet transformation rescales the lens-plane coordinates and inferred luminosity distance (Eq.~\ref{eq:F_mass_sheet}), a different magnification affects the mass and offset of the microlens and the source's redshift and intrinsic component masses (App.~\ref{app:source_properties}, Fig.~\ref{fig:magnification_vs_mass}).
GW231123 provides a particularly informative case study for multi-scale inference, and demonstrates that neglecting these effects can bias source parameter estimation, with downstream implications for the inferred formation channel.


\section{Bayesian Inference}

We performed Bayesian parameter estimation on GW231123 under three hypotheses: \textit{unlensed}, lensing by \textit{isolated} point mass lens (PL) and lensing by  \textit{embedded} PL, as described above. We followed the same configuration as Ref.~\citep{LIGOScientific:2025rsn} but without distance marginalisation, while interfacing the most recent version of \textsc{Bilby} (version 2.6) \citep{bilby_paper,Romero-Shaw:2020owr} with a modified version of the \glow{} code~\citep{Villarrubia-Rojo:2024xcj} for computing lensed waveforms.  We used publicly available time-series data for LIGO Hanford (LHO) after glitch subtraction and LIGO Livingston (LLO) from GWOSC\footnote{\url{https://gwosc.org/eventapi/html/O4_Discovery_Papers/GW231123_135430/v1/}}, along with noise power spectral density and detector calibration envelopes from the data released on Zenodo \citep{zenodo}.

For each hypothesis, we considered \textit{three} BBH source waveform approximants:
\texttt{IMRPhenomXPHM-ST} (Phenom) \citep{Colleoni:2024knd,xphm} and \texttt{NRSur7dq4} (NRSur) \citep{Varma:2019csw} and \texttt{SEOBNRv5PHM} (SEOBNR) \citep{Ramos-Buades:2023ehm}. Our unlensed hypothesis results are consistent with Ref.~\citep{LIGOScientific:2025rsn} (see Table \ref{table:source}, in App. \ref{app:source_properties}). 
The isolated PL hypothesis has \textit{two} extra parameters, redshifted lens mass $M_{Lz}=M_L(1+z_L)$ and impact parameter $y$.  The Embedded PL has \textit{four} extra parameters,  rescaled external shear $\tilde \gamma$, redshifted lens mass $M_{Lz}$, and microlens position ($\tilde y$, $\theta_L$), as compared to the unlensed hypothesis.  The priors for the BBH source parameters are kept the same as \cite{LIGOScientific:2025rsn} for all hypotheses, and $M_{Lz}$ logarithmic uniform in $[10, 10^5]$, $y$ and $\tilde y$ uniform in $[0.05, 5]$. Additionally, for embedded PL we set uniform priors for  $\theta_L$ in $[0, \pi/2]$ and $\tilde \gamma$ in $[0, 1]$. These choices are intended as agnostic, rather than astrophysically-informed. Our results are summarised in Table ~\ref{table:lens}.

\begin{table}[t!]
\centering
\renewcommand{\arraystretch}{1.5}
\begin{tabular}{l|ccc}
\hline
\hline

 & NRSur & Phenom & SEOBNR \\
\hline
$\log_{10} \mathcal{B}^{L}_{U}$ &$2.82$ & $5.0$& $1.24$\\
&($2.91)$ & ($4.68$)& ($1.13$)\\
\hline
$\Delta \log \mathcal{L}_{\rm max}$ &$6.3$ &$4.1$ & $5.8$\\
&($4.8$) &($1.9$) & ($2.9$)\\
\hline

\hline \hline
$M_L (1+z_L)$ $[M_\odot]$ & $672^{+553}_{-363}$ & $558^{+675}_{-277}$ & $576^{+1012}_{-344}$ \\
&  ($821^{+689}_{-322}$) & ($942^{+730}_{-325}$) & ($818^{+723}_{-493}$) \\

\hline
$\theta_L$ & $0.5^{+0.51}_{-0.41}$ & $0.37^{+0.41}_{-0.31}$ & $0.43^{+0.73}_{-0.38}$ \\
\hline

$\tilde{\gamma}$ & $0.66^{+0.2}_{-0.61}$ & $0.73^{+0.17}_{-0.58}$ & $0.49^{+0.35}_{-0.45}$ \\
\hline
 $\tilde{y}$ & $1.5^{+1.7}_{-0.88}$ & $1.8^{+2}_{-0.97}$ & $1.3^{+2.3}_{-0.92}$ \\
\hline\hline
$y$ & $1.9^{+2.4}_{-1.2}$ & $2.4^{+2.9}_{-1.4}$ & $1.6^{+2.9}_{-1.2}$ \\
& ($0.68^{+0.39}_{-0.3}$) & ($0.59^{+0.29}_{-0.25}$) & ($0.67^{+0.83}_{-0.31}$) \\
\hline
$\kappa = \gamma$ & $0.4^{+0.07}_{-0.35}$ & $0.42^{+0.05}_{-0.29}$ & $0.33^{+0.13}_{-0.29}$ \\
\hline
$\mu_{\rm macro}$ & $5^{+9.2}_{-3.8}$ & $6.5^{+14}_{-5.2}$ & $2.9^{+8.2}_{-1.8}$ \\
\hline
$\ymacro$ & $0.25^{+8.4}_{-0.18}$ & $0.18^{+2.5}_{-0.13}$ & $0.52^{+11}_{-0.42}$ \\
\hline
$P_{\rm SL}$ & $75.3 \%$  & $88.4 \%$ & $62\%$ \\
\hline
\hline
\end{tabular}
\caption{Summary of results under \textit{lensing by an embedded (isolated) point mass} hypothesis. Maximum likelihood is relative to unlensed NRSur. See Table \ref{table:source} for the recovered source parameters and Fig. \ref{fig:time_corner} for some 2D posteriors of the lens parameters.
\label{table:lens}}
\end{table} 

We find evidence for lensing, the Bayes factors for lensed v/s unlensed hypothesis being  $\log_{10}\mathcal B^{L}_{U} = 2.82 \left(2.91\right)$, $5.0 \left(4.68\right)$ 
and $1.24 \left(1.13\right)$  for embedded (isolated) PL using NRSur, Phenom and SEOBNR waveforms, respectively. Note that for the isolated PL case, though our results are similar to those of Ref. \citep{lvklensing}, our Bayes factors are not the same because of different prior choices. The embedded PL gives a higher (lower) Bayes factor than isolated PL for Phenom and SEOBNR (NRSur), but invariably gives a better fit (see Table \ref{table:lens}).%
\footnote{Ref.~\citep{LIGOScientific:2025rsn} also considers waveform approximant  \texttt{IMRPhenomXO4a}, finding it to fit the GW231123 data better than the rest of the approximants under the unlensed hypothesis. Since it is less reliable in the highly spinning regime, we don't consider it here and point the reader to \cite{lvklensing} for isolated PL case.} 
This is expected, as each model is a direct extension of the previous one. 
The large variation in the Bayes factors across waveforms 
stems from model discrepancies when both the binary components are highly spinning (both components spin magnitudes $>0.8$~\cite[Fig.~2]{LIGOScientific:2025pvj}), as preferred in the unlensed analysis.
 
Only the NRsur waveform achieves the required accuracy in 90\% of cases, when compared to the latest NR simulations~\citep{LIGOScientific:2025rsn}. Therefore, we will now quote results using NRSur in the main text. While very high, we note that these Bayes factors need to be interpreted with caution before claiming a decisive lensing detection: injection studies are needed to determine whether detector noise can mimic diffraction signatures \citep{LIGOScientific:2023bwz, mishrapopbiases2023}, especially as the GW signal is of very short duration.

We now estimate the significance of a lensing detection in GW231123 with an injection study, assuming Gaussian detector noise. Because Bayes factors are sensitive to prior choices and noise fluctuations, the false alarm probability (FAP) needs to be estimated from the simulated background distribution of $B_U^L$.

\begin{figure}
    \centering
    \includegraphics[width=\linewidth]{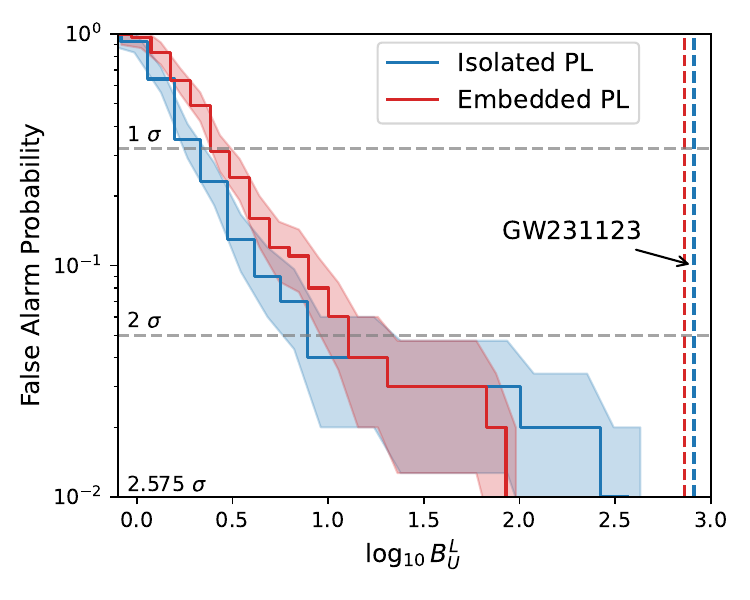}
    \caption{False alarm probability of observing Bayes Factors from GW231123-like unlensed signals. The curves are generated by analysing a hundred unlensed NRsur signals injected into Gaussian noise with lensed and unlensed hypotheses. The shaded region shows the Poisson error. Horizontal lines show the corresponding confidence levels. The GW231123 Bayes factor lies outside the distribution, with FAP $<1\%$ and significance $\gtrsim 2.6 \sigma$.}
    \label{fig:background}
\end{figure}

We inject one hundred unlensed signals with parameters drawn from the GW231123 unlensed NRSur posteriors into simulated Gaussian noise realisations generated using the same power spectral density as for the real event. We then perform parameter estimation on these signals under the unlensed, isolated PL, and embedded PL hypotheses and compute the Bayes factors. 
Fig.~\ref{fig:background} shows the false alarm probability, which is the same as the inverse cumulative density function of Bayes factors from unlensed signals.
The signal's Bayes factors under both embedded and isolated PL hypotheses are found to be higher than any of the injections, allowing us to establish a FAP$<1\%$ of the lensing conclusion, 
corresponding to $\gtrsim 2.6 \sigma$ confidence level. Note that the Bayes factors may change if the data contains micro-glitches or non-Gaussian noise~\citep{Ray:2025rtt,Bini:2026kwz}. \cite{lvklensing} constructed a background assuming a population model of BBH sources, injected into real noise, finding FAP $<0.39\%$ for isolated PL hypothesis.

\begin{figure*}
    \centering
    \includegraphics[trim=20 15 0 0, clip,width=0.7\linewidth]{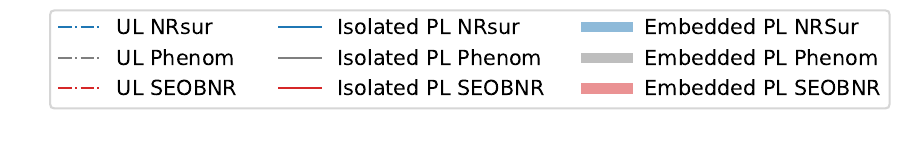}

    \includegraphics[width=0.45\linewidth]{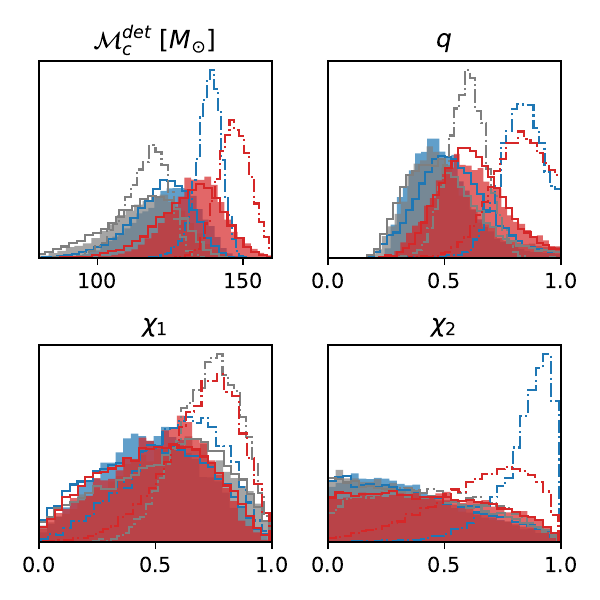} \hfill
    \includegraphics[width= 0.45\linewidth]{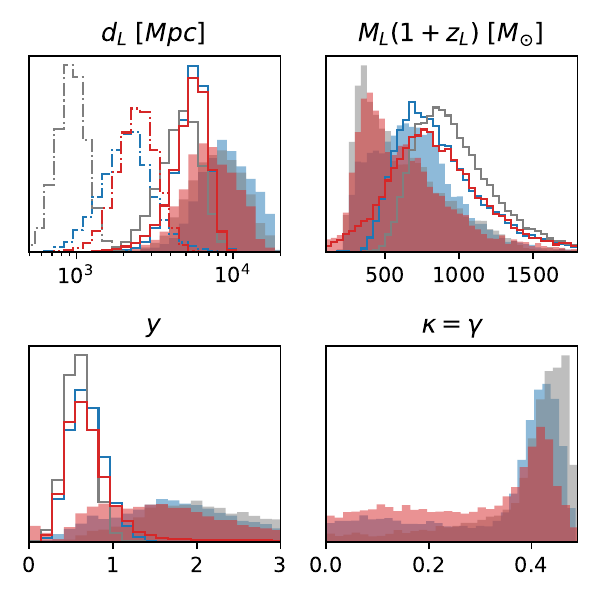}\\

    $(i)$ Waveform systematics and source parameters \hspace{ 0.1 \linewidth} $(ii)$ Microlens parameters and luminosity distance\\
    
    \caption{ 1D marginalised posterior distributions of $(i)$ detector frame chirp mass $\mathcal{M}_c^{\rm det}$, mass ratio $q$ and projected spin components $\chi_1$ and $\chi_2$ (perpendicular to the binary orbital plane) under unlensed and lensed hypotheses with different waveforms. The posteriors are in agreement for all the waveform models after incorporating lensing by an isolated or embedded point mass. $(ii)$ luminosity distance of the source $d_L$, and parameters of the point mass lens (redshifted lens mass $M_L(1+z_L)$, impact parameter $y$) without and with external potential (assuming $\kappa = \gamma$). The posteriors under both lensing hypotheses are consistent among the three waveforms. 
    \label{fig:1D_systematics_microlens}}
\end{figure*}

The lensing analysis can be interpreted as the superposition of a main signal and an additional feature, delayed by $\sim 20$ ms. The Embedded PL posteriors display multi-modality, and can be decomposed into four distinct regions: 
1) two microimages $\Delta t\sim 22$ ms, 2) two microimages $\Delta t\sim 44$ ms, 3) two microimages $\Delta t\sim 66$ ms, 4) four 
microimages. The identification of the posterior regions is shown in Fig. \ref{fig:time_corner} (App.~\ref{app:timed_domain}). None of the modes can be fully accounted for with geometrical-optics lensing, as they have a common diffractive wave-optics feature at $\sim 20$ ms (see Fig. \ref{fig:time_bestfit}, App.~\ref{app:timed_domain}), responsible for the better fit to the data. Hence, diffractive/wave-optics signatures are essential to our physical modeling of GW231123 as a lensed source. 
A more comprehensive lens model (e.g.~including the effects of many light stars) may provide additional evidence for lensing.

\section{Source and Lens Properties} \label{sec:lens_source_prop}
Here we discuss the inferred properties of the BBH source, the microlens, and the possibility of observing a second macroimage. We will quote results only with NRSur waveform, which are qualitatively similar to the other two (see Tables \ref{table:lens} and \ref{table:source}). We also comment on the astrophysical implications of our findings, but leave detailed estimates for future work.

\subsection{Source properties and waveform systematics} 

The lensing hypothesis is not only favoured, but also removes many unusual features of GW231123.
First of all, the results of the three waveform approximants are consistent under the lensing hypothesis ({see also \cite{lvklensing}), for both source and lens
parameters, unlike the unlensed analysis, where there are discrepancies among the waveform approximants, expected due to very high component spins ~\citep{LIGOScientific:2025rsn}. This could be because the lensed hypothesis brings the source parameters into the regime where the waveforms are accurate and agree with each other, making the posteriors consistent across waveform models. 
Fig. \ref{fig:1D_systematics_microlens} shows 1D marginalised distributions for all waveform models. The posteriors in detector frame chirp mass $\mathcal{M}_c^{\rm det}$, mass ratio $q$, and projected spin components  $\chi_1$ and $\chi_2$ (perpendicular to the orbital plane) are in agreement for the three waveforms for both isolated and embedded PL. Support for spin precession is also lower in the isolated/embedded PL analyses $\chi_p \sim 0.5$ as compared to unlensed $\chi_p \sim 0.75$, as expected from the correlation between precession and diffraction~\citep{Shan:2025jpt}. Additional details on the source parameters are given in App.~\ref{app:source_properties}. 

The lensing hypothesis aligns GW231123 with the typical source masses and distances of previously found events. 
Fig.~\ref{fig:mass_redshift_contours} shows how the combination of magnification and redshift drives the posterior towards GWTC-4 events, both for the isolated and embedded PL. 
For the embedded PL, including the external convergence by assuming $\kappa=\gamma$ (SIS macrolens prescription) translates to a farther source with median redshift $z\sim 1.2$ compared to $\sim 0.38$ (magnified by $\mu_{\rm macro}~\sim 5$) and $16\%$ of the posterior at redshifts beyond the nominal detector horizon. The additional redshift makes the binary intrinsically less massive with median total source mass $M^{\rm src}_{\rm tot} \sim 137 M_\odot$ compared to $\sim 232 M_\odot$ and median binary source component masses below $100$ and $50M_\odot$ (see table \ref{table:source}). 
Note that unlensed, isolated PL and embedded PL are nested models, each one a 2-parameter extension of the former. 
The broadening and shift of the posteriors arise not only from the additional degrees of freedom in the lensing models, but also because the unlensed analyses overconstrain and bias the region preferred by the data.

Under the unlensed hypothesis, GW231123 can be explained only as a rare draw from the population, due to a drop in merger rates for high component mass, with both black holes $>100 M_\odot$ \citep{LIGOScientific:2025pvj}. 
It also challenges our understanding of binary formation, as at least one of the binary components falls in the mass gap 60\text{--}130 $M_\odot$, where pair-instability prevents the black hole formation by stellar collapse \citep{pisn}. 
Studies have also shown that hierarchical mergers or formation in active galactic nuclei are unlikely \citep{passenger2025gw231123hierarchicalmerger,delfavero2025prospectsformationgw231123agn}, requiring it to be mergers of at least second-generation black-holes. The inclusion of GW231123 as unlensed is likely to affect population inference: a peak at $z\sim 0.2$, $\Msrctot\sim 250\, M_\odot$ is visible in model-agnostic inference~\cite[Fig.~2]{Tenorio:2025nyt}.

The lensing hypothesis makes the source less extreme from an astrophysical perspective: a higher redshift increases the comoving volume and overall merger rate, while lower source masses bring the event closer to the high-mass end of the BBH population.
With lensing, the median secondary mass lies below the mass gap, and the primary can lie within the mass gap. Since lensing is also a rare phenomenon $P(\mu>2) \sim 0.01-1\%$ \citep{Oguri:2018muv}, we expect the prior probabilities of both hypotheses to be low. In principle, the prior probabilities would also depend on the abundance of the microlens of masses 100--1000 $M_\odot$, which is largely unknown. A detailed analysis is left for the future. %

The achromatic macro-magnification is not fixed by the waveform: a more strongly magnified macro-configuration ---e.g.~near the caustics of a realistic elliptical lens--- lowers the source-frame masses and can bring the primary below the pair-instability gap, while also reducing the physical mass required of the microlens (see App.~\ref{app:source_properties} and Fig.~\ref{fig:magnification_vs_mass}).

Although highly magnified events are rare, a rate of $\sim 1$ event/yr with $\mu>2$ is consistent with the upper edge of estimates at O4 sensitivity, which corresponds to $\sim 10$ events/yr at O5 sensitivity~\cite[Tab.~3]{Smith:2022vbp}, \cite{aplus}. 
These results assume that the BBH merger rate is proportional to star formation; the number may be higher if the merger rate increases more steeply than star formation~\citep{Mukherjee:2021qam} or in the presence of a high-redshift population~\citep{Ng:2020qpk}.

\begin{figure}[t]
    \centering
    \includegraphics[width=1\linewidth]{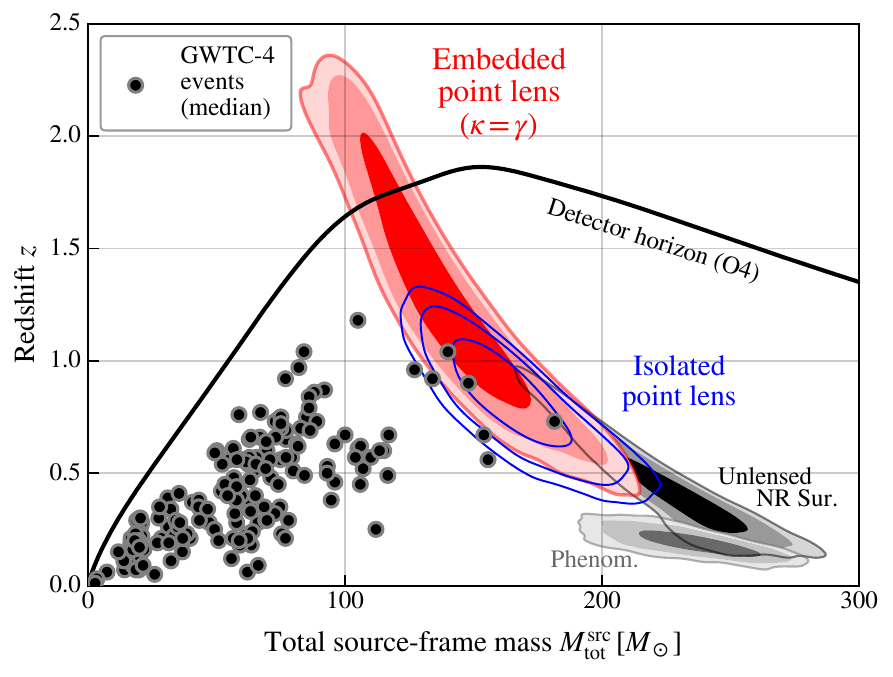}
    \caption{GW231123 total source-frame mass $M^{\rm src}_{\rm tot}$ versus redshift $z$, as a lensed black hole merger.  
    Contours indicate 68, 95, and 99\% credible regions for different waveform and lensing models: unlensed (gray filled; Phenom and NRSur), diffracted (blue unfilled; NRSur), and magnified+diffracted by a singular isothermal macrolens (red filled; NRSur). Circles indicate median values of GWTC-4 events \citep{gwtc4catalog}.
    The black curve marks the approximate detector horizon for O4.  
    Unlike the unlensed interpretation, which requires unusually high masses relative to the GWTC-4 population \citep{gwtc4pop}, lensing analyses harmonize the event with the broader distribution of binary black holes. }
    \label{fig:mass_redshift_contours}
\end{figure}

\subsection{Microlens properties} 

Our analysis involves a single compact microlens, whose redshifted mass is constrained $M_L(1+z_L) \sim 300-1500 M_\odot$, with the embedded PL requiring masses that are $20-40$\% lower than the isolated PL (cf.~Tab.~\ref{table:lens}, Fig.~\ref{fig:1D_systematics_microlens}, right). 
The intrinsic mass $M_L$ is also lower due to the lens redshift: the embedded PL has intrinsic masses $M_L\in (188,850) M_\odot$ while the isolated PL yields $M_L\in (468,1184) M_\odot$ at 90\% c.l., where a flat mass function is assumed for the microlenses (see App.~\ref{app:microlens_mass}).%
\footnote{The inferred quantity is the effective lensing mass scale controlling the wave-optics response; for extended lenses, this scale can be orders of magnitude lower than the total physical mass of the object, e.g.~\cite[Eq.32]{Tambalo:2022wlm}.}
The impact parameter $y$ is a factor $2.4-4$ times higher for embedded PL ($\kappa=\gamma)$ than in the isolated case, allowing a much larger offset between the (projected) source position and the microlens.  
The number of microlenses at an impact parameter $\leq y$ scales with the area as $\propto y^2$: this factor alone increases the probabilities of a chance alignment by an order-of-magnitude in the embedded case.

A simple possibility is that the microlens is an astrophysical compact object, such as an intermediate-mass black hole. This interpretation is not unique, and should be treated with caution given the uncertain abundance and observational status of such objects.
In this case, the embedded PL emerges as more natural, as these black holes will be predominantly located in galaxies, which provide the external potential $\kappa,\gamma\neq0$. In this case, the lower $M_L$ makes their abundance more plausible than in the isolated case, as heavier objects are more rare~\citep{Sicilia:2021gtu}. 
However, a negligible external potential is expected if the lens and source are in close proximity, e.g.~both the lens and source belong to the same stellar cluster~\citep{Ubach:2025dob}. 
This scenario can be further distinguished by strong-gravity effects, peculiar binary acceleration~\citep{Oancea:2022szu,Samsing:2024xlo} and if $\gamma,\kappa\sim 0$.
Finally, it is possible that the microlens represents an extended object, e.g.~the center of a globular cluster: $M_{Lz}$ is then the redshifted mass projected within the Einstein radius~\cite[App.~B]{Savastano:2023spl}.
This possibility could be established considering diffraction by an extended lens~\citep{Cheung:2024ugg}.

A more speculative possibility is that the effective microlens is associated with a dark-matter structure compact enough to produce point-lens-like diffraction, e.g.~
in theories that predict objects with a range of masses and profiles~\citep{Zumalacarregui:2024ocb,Croon:2024jhd,Bringmann:2025cht,Cheung:2026pky}.
Assuming that unlensed events can probe similar objects up to ${2\times} \,  (1\times)$ their Einstein radius~\citep{mishrapopbiases2023}, the fraction of dark matter in compact objects needed to explain GW231123 is $f_{\rm c} \equiv \frac{\Omega_{\rm c}}{\Omega_{\rm CDM}} > 5.2\, (12.2)\%$ for the embedded PL and $f_{\rm c}> 6.4\, (20.9\%)$ for the isolated case, at 90\% c.l. (see Fig.~\ref{fig:dark_matter_after_GW231123}). Even for extended or late-forming objects, these values are excluded by stellar lensing through the Milky Way halo, stability of wide binaries and stellar dynamics in ultra-faint dwarf galaxies~\citep{Mroz:2024mse,Shariat:2025dxs,Graham:2025opw}. See App.~\ref{app:microlens_abundance} for derivation and discussion of these results.
 
One has to be mindful of the limitation of assuming a single microlens. It is conceivable that more complex/realistic microlens distributions, containing many objects, can mimic the wave-optics lens features seen in GW231123. Even if an intermediate-mass object dominates the signal, additional signatures caused by a large number of light stars may help confirm or disprove the lensing hypothesis.
We show this explicitly in App.~\ref{app:microlens_alternatives}: both an extended object with $M_{\rm vir}\sim 10^4-10^6 M_\odot$ and a collection of $\sim 10M_\odot$ objects can closely reproduce the wave-optics signature of a single heavy point lens (Fig.~\ref{fig:microlens_alternatives}).
Thus, while the data constrain the effective lensing mass scale, identifying the physical nature of the microlens requires additional lens modeling and astrophysical priors \citep[e.g.][]{Cheung:2026pky}, as discussed in App.~\ref{app:source_properties}.


\subsection{Additional macroimages}

Assuming a macrolens profile enables concrete predictions for the existence and properties of additional images (i.e. strong lensing). 
In the SIS model, the offset between the source and the macrolens' centre (Fig.~\ref{fig:lensing_diagram})  is related to its convergence, $y_{\rm macro} =\frac{1}{2\kappa}-1$. The median $\kappa=\gamma \sim 0.4$ suggests that the lens is located in the bulk of an embedding macrolens (galaxy) with median $y_{\rm macro}\sim 0.25$.
Our analysis of the embedded PL analysis gives a strong-lensing probability $P_{\rm SL} = P(\ymacro < 1) \sim 75\%$, associated to the formation of a second macroimage after GW231123, with a delay
\begin{equation}
    \Delta t_{\rm macro} \simeq 3.4_{-3.4}^{+9.9} \,\mathrm{days} \left(\frac{\ymacro}{0.25} \right)\left(\frac{M^{\rm eff}_{\rm macro}}{10^{11}\,M_\odot} \right) \,,\label{eq:macro_delay}
\end{equation}
for $\ymacro\leq 1$.
Here $M^{\rm eff}_{\rm macro}$ is the projected macrolens mass enclosed within its Einstein radius, and the reference value corresponds to a typical galaxy (total mass $10^{12}M_\odot$~\citep{Robertson:2020mfh}) and the coefficient amounts to marginalizing over $(1+z_L)D_L/D_{LS}$ from the inferred lens redshift (App.~\ref{app:microlens_mass}).
The strong lensing scenario is compatible with expectations for multiple detectable images at O4 sensitivity \citep{o4}, with rate $\sim 0.65$/yr~\cite[L/H Tab.~1]{Wierda:2021upe}.

Assuming that the observed event corresponds to a type I macroimage produced by an SIS, we can predict the relative amplitude of 
the secondary type II macroimage as $R^{II}_{I} \equiv \sqrt{\frac{|\mu^{II}_{\rm macro}|}{|\mu^I_{\rm macro}|}} = \left|\frac{2} {\mu_{\rm macro}^I} -1 \right|^{1/2}$, with a median value $R^{II}_I \sim 0.82$. 
Given the signal-to-noise ratio (SNR) of GW231123, ${\rm SNR}_I\sim 20$, the second image can be loud ${\rm SNR}_{II}\sim R^{II}_{I} {\rm SNR}_I\sim 16$. However, the actual SNR is a function of $\Delta t_{\rm macro}$ through variations in detector sensitivity (noise, Earth's rotation modulation of antenna pattern), so we find the probability of detectable second image to be ${P_{\rm SL}(\rm SNR}_{II}>8) \sim 55 \%$.

\cite{lvklensing} searched for possible strongly lensed counterparts of GW231123 using the unlensed posteriors, finding no significant candidates. 
The lack of reported events consistent with GW231123 does not necessarily rule out the strong-lensing hypothesis: Additional images should be consistent with the intrinsic parameters of the diffracted source (PL or Embedded PL, Fig. \ref{fig:1D_systematics_microlens}) and the much wider sky localization of the lensed analyses (Fig. \ref{fig:sky} in Supplemental Material).
The second image will form closer to the center of the lens, increasing diffraction distortions (larger shear and microlens density) and the chance of it being missed when comparing source parameters~\citep{mishrapopbiases2023} or by GW searches based on unlensed templates~\citep{Chan:2024qmb}.
The lack of additional images can also be explained by data gaps or the event becoming sub-threshold (e.g.~due to antenna-pattern modulations or low $R_{I}^{II}$).  
Finally, it is plausible that the macrolens is a galaxy group or a cluster, for which time delays (Eq.~\ref{eq:macro_delay}) can be years and the relationship between magnification, shear and image multiplicity is much more complex~\citep{Vujeva:2025kko}.
Additional images of GW231123, if found, will provide further evidence for a lensing detection and give information on the macrolens and source properties.


\section{Conclusions and prospects} \label{sec:concl}

Standard GW231123 analysis requires a fairly close ($z<0.3$) source with total mass and component spins beyond previously identified signals, and leads to unprecedented discrepancies among waveform approximants. These results challenge our understanding of the formation of this binary system. We find evidence consistent with gravitational-wave lensing at a false-alarm probability of $\lesssim 1\%$, representing a conservative upper bound set by the number of simulated background realizations (Fig.~\ref{fig:background}).
Lensing by an isolated point lens (PL) brings the waveform analyses into closer agreement and shifts the source toward less extreme masses and spins, but also requires the lens to be a $\sim 1000 M_\odot$ object closely aligned with the source.
The same diffraction pattern can, however, be produced by astrophysically ordinary configurations: a more massive extended lens (e.g. a globular cluster) or a dense collection of stellar-mass objects (App.~\ref{app:microlens_alternatives}).

Our lensing analysis incorporates an external gravitational field, which is expected around microlenses embedded in a massive object (macrolens).
Most of the parameters are recovered, including source properties, two components of the external potential and the microlens mass, which is lower than for the isolated PL. 
Wave-optics signatures probe propagation effects directly encoded in the waveform (microlens mass, alignment, external potential). This is in contrast with alternative scenarios (e.g. AGN, hierarchical, population III, primordial), where the system's properties are inferred indirectly from masses and spins.

The analysis is limited by a fundamental ambiguity (mass-sheet degeneracy), which can be broken by assuming a relation between the isotropic and anisotropic components of the external potential. Choosing a macrolens profile inspired by a symmetric galaxy (singular isothermal sphere) fully determines the gravitational magnification and the source's true redshift, which may lie beyond the nominal detector horizon. 
The galactic assumption also increases the physical offset between the microlens and the projected source position, increasing the probability of a chance alignment by an order of magnitude, relative to the isolated PL results. Combining the external potential with a macrolens model allows us to predict the properties of additional images, whose discovery may confirm the lensing hypothesis.


Although our analysis employs simplifications, it illustrates how lensing signatures provide valuable information about distant GW sources and gravitational lenses, from stellar to galactic scales. 
Some of our prescriptions, such as the macroimage/lens properties, can be easily generalized to account for realistic galaxies, groups, and clusters, as informed by observations and simulations.
With respect to lensing diffraction, each analysis is a direct extension of the former: the embedded PL supersedes the isolated one, which itself generalizes the unlensed analysis.
Quite remarkably, each subsequent layer of detail is not only constrained by the data, but also provides a better fit, new insights, and a more plausible astrophysical interpretation of the source-lens system.

A fundamental limitation involves the challenges of wave-optics computation and the need to describe microlensing by complex matter distributions. Even with the latest developments in wave-optics algorithms, our analysis was possible only for a rather simple lensing potential. 
In contrast, realistic stellar fields are described by thousands of microlens masses and positions, which cannot be explored systematically. 
A theory of lens stochastic diffraction and model-agnostic analyses need to be developed to confidently discover and interpret lensed GWs. 
A better understanding of GW emission by highly-spinning systems is needed to better assess the unlensed hypothesis.

Additional lensing signatures may confirm the lensing scenario at high confidence and allow us to answer the most intriguing questions around GW231123, including its formation channel, the distance to the source and the nature of the microlens.
Full Bayesian parameter estimation with wave-optics diffraction from realistic stellar fields or stochastic compact-object distributions remains beyond the current state of the art, although effective descriptions of stellar-field diffraction are now becoming available~\citep{Zumalacarregui:2026uqs}; the embedded point-lens model studied here provides a controlled description of the leading effect of a dominant compact object in an external potential, highlighting the need for more work in this direction.

Independently of this particular event, our work demonstrates how current GW observations already probe a qualitatively new regime, where wave-optics lensing distortions link stellar-mass objects ($\lesssim 10^3 M_\odot$) to macroscopic lenses ($10^{11}M_\odot$).
As GW detectors improve in sensitivity and event rates increase, this multi-scale lensing framework opens a clear path toward identifying and exploiting lensing signatures as a new probe of compact objects, galaxy-scale structure, and fundamental physics.


\begin{acknowledgements}
        We are very grateful to Alessandra Buonanno, Juan Calderon Bustillo, Mark Cheung, 
        Djuna Croon, Liang Dai, Arnab Dhani, Chema Diego, Pierre Fleury, 
        Jonathan Gair, Alice Garoffolo, Serena Giardino, Justin Janquart, Nicola Menadeo, Anuj Mishra, Xikai Shan, Sherry Suyu, Alex Toubiana, Jay Wadekar, Jenny Wagner
        Yifan Wang and Matias Zaldarriaga for valuable discussions.
   M.Z. acknowledges financial assistance from the Max Planck Society Supporting Members, which was crucial to the completion of this work, and support from the European Research Council (ERC) under the European Union’s Horizon Europe research and innovation programme (ERC CoG 2025 – GLOW, Grant agreement No. 101230608).
    Besides the libraries mentioned in the main text, \glow{} also relies on Numpy~\citep{harris2020array}, Scipy~\citep{2020SciPy-NMeth} and Colossus~\citep{Diemer:2017bwl}. Most of the analysis was performed on hypatia, an AEI cluster.
 \\ \indent  This material is based upon work supported by NSF's LIGO Laboratory which is a major facility fully funded by the National Science Foundation,  and has made use of data, software and web tools obtained from the Gravitational Wave Open Science Centre (\url{www.gw-openscience.org}). The authors gratefully acknowledge the LVK Collaboration for providing the publicly released GWTC-4 data.
Large language models (Anthropic's Claude, OpenAI's ChatGPT) were used as assistive tools in this work; the authors verified all results and are fully responsible for the content.

\end{acknowledgements}

\section*{Data Availability}
The GW strain data used in this work are publicly available from the Gravitational Wave Open Science Center (GWOSC) for event GW231123. Noise power spectral densities and calibration information were obtained from the associated public release. Posterior samples and derived data products generated in this study will be made publicly available upon publication through a Zenodo record with reserved DOI \href{https://doi.org/10.5281/zenodo.20217942}{10.5281/zenodo.20217942}.
Scripts that reproduce figures and derived numbers from this work are publicly available at \url{https://github.com/miguelzuma/GW231123_lensing_PLS}.

\clearpage




\setcounter{equation}{0}
\setcounter{figure}{0}
\setcounter{table}{0}
\setcounter{section}{0}

\newcommand\ptwiddle[1]{\mathord{\mathop{#1}\limits^{\scriptscriptstyle(\sim)}}}

\appendix

\makeatletter
\@removefromreset{equation}{section}
\makeatother

\renewcommand{\theequation}{S\arabic{equation}}
\renewcommand{\thefigure}{S\arabic{figure}}
\renewcommand{\thetable}{S\arabic{table}}



\twocolumngrid


\section{Microlensing diffraction analysis}\label{app:timed_domain}

One important feature of our results is a very clear multimodality in the posteriors, see Fig. \ref{fig:time_corner}. We observed
this multimodality in the lens parameters for all the waveform generators in the Embedded PL. In order to understand its origin, it
is convenient to work with the time-domain version, $I(\tau)$,  of the amplification factor, $F(w)$, 
\begin{equation}
    F(w) = \frac{w}{2\pi\ii}\int^\infty_{-\infty} \di\tau\ \ee^{\ii w\tau}I(\tau)\,.
\end{equation}
Using this quantity, it is easier to interpret the results and highlight the importance of wave-optics effects. Geometric optics
effects appear in $I(\tau)$ as discontinuities, either singular peaks for Type II images or step functions for Type I and III images.
The relevance of wave-optics effects can then be addressed by subtracting this (singular) geometric optics contribution from the full
$I(\tau)$. The resulting (regularized) $I(\tau)$ can still present peaks, or diffractive features, with very important observable
consequences. 
We emphasize that $I(\tau)$ is a lensing time-delay response, not the detector-frame strain. A localized feature in $I(\tau)$ generally produces frequency-dependent modulation of the waveform through $F(f)$, and therefore need not appear as a residual confined to a narrow interval in detector time.

Fig. \ref{fig:time_bestfit}, contains the best-fit results for the three main modes identified in our analysis. Here we can see that the full amplification factor is the result of an interference between a Type I image (the main image), a very important diffractive feature (smooth peaks in the figure) and a secondary Type II image (sharp peaks in the figure). Moreover, the main contributing factor to the overall shape of $F(f)$ is the interference between the diffractive feature and the main image~\citep{Brando:2024inp,Jow:2022pux,Meena:2025gry}. 
The posterior modes are sharply localized in lensing delay, with microimage delays around $\Delta t_{01}\sim 22,44,$ and $66\,{\rm ms}$, and all contain a prominent structure near $20\text{--}22\,{\rm ms}$ in the lensing response.
The multimodality in the posteriors arises from the fact that multiple physical configurations can achieve roughly the same diffractive feature.

Fig.~\ref{fig:231123_lensed_time} shows the time domain waveform for GW231123. The top panel includes the best fit for embedded PL and unlensed, along with a model-independent reconstruction produced by the coherent wave burst -- CWB pipeline \citep{Klimenko:2008fu}. The bottom panel shows the decomposition as a microlensed event: the best-fit waveform is split into main image, secondary image and diffraction feature. The best-fit signal has a time delay $\Delta t \sim 44$ms between the main and secondary microimages, corresponding to the blue contours in Fig.~\ref{fig:time_corner}, and the blue line in Fig.~\ref{fig:time_bestfit}.

It is important to emphasize that the parameters of the embedded microlens system are inferred from features in the waveform, rather than imposed by the macrolens model.
Fig.~\ref{fig:pls_parameters_wo} illustrates how the amplification in the time and frequency domain responds to variations of the external potential magnitude $\gamma$ and relative orientation $\theta_L$ ($\gamma=\kappa$ is assumed).
In the absence of microlens, $\gamma,\kappa$ would only rescale the signal amplitude, introducing no frequency dependence and remaining fully degenerate with luminosity distance.
In the presence of a microlens, however, the same external potential changes the local microcaustic structure and the time-delay distribution of microimages and diffractive features.
As a result, varying either the strength of the external potential or the position of the microlens produces distinct, frequency-dependent modulations in $F(f)$.
The macro-potential therefore becomes observable through its effect on the chromatic wave-optics response of the embedded microlens.
These distortions are chromatic precisely because they originate in wave-optics diffraction: the amplification factor depends on the ratio of the GW wavelength to the time-delay scale of the lens. In contrast, geometric-optics (macro)lensing produces achromatic magnification, degenerate with the luminosity distance.
This is reflected in the posterior distributions, where the microlens mass, impact parameter, position angle, and reduced shear are all constrained by the data, cf. Fig.~\ref{fig:time_corner}.

    \begin{figure*}
        \centering
        \includegraphics[width=0.99\linewidth]{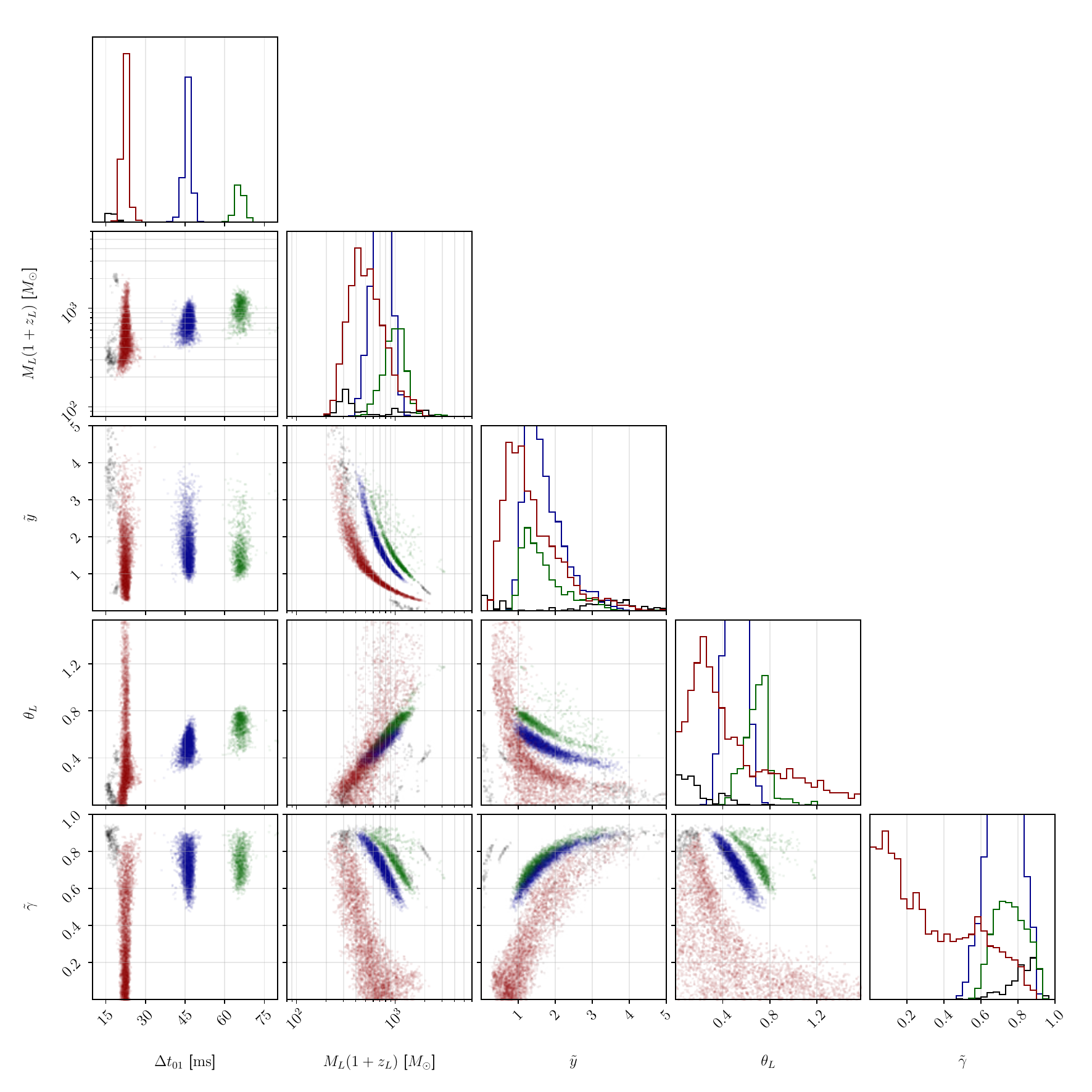}
        \caption{Different microlensing regions inferred from the Embedded PL (NRSur) analysis, showing the multi-modality in the posteriors, related to the properties of the microimages. We identified a very subdominant mode with 4 images (black, $4\%$ of samples) and three main modes with 2 images, which can be classified in terms of the time delay between them $\Delta t_{01}$: $\sim 22$ ms (red, $43\%$), $\sim 44$ ms (blue, $38\%$) and $\sim 66$ ms (green, $15\%$). The best fit is provided by the mode at $\sim 44$ ms (blue), see Fig. \ref{fig:time_bestfit} for
        the best fitting results for each of the modes.}
        \label{fig:time_corner}
    \end{figure*}

    \begin{figure*}
        \centering
        \begin{minipage}{\textwidth}
            \centering
            \includegraphics[width=0.99\linewidth]{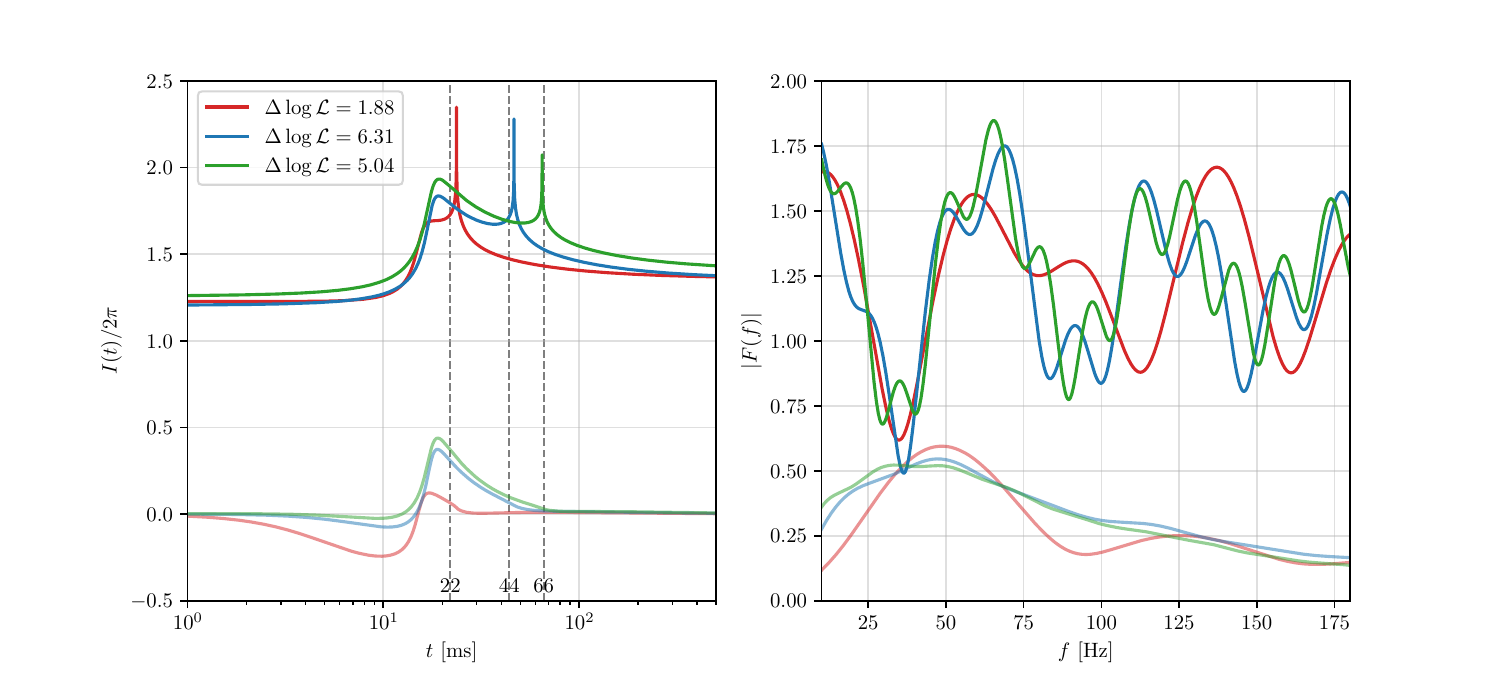}
            \caption{Amplification factor with the best-fit lensing parameters for the three modes with two images shown in
            Fig. \ref{fig:time_corner}. The reference likelihood value is the unlensed case. The wave-optics contribution (i.e. full
            result minus the geometric optics contribution) is represented with a lighter shade.
            In all three cases, the diffractive feature at $t\sim 20$ ms (left) is very important (note that the feature is defined in lensing time delay, not detector time). The interference between the diffractive feature and the main image produces the main
            oscillatory pattern observed in $F(f)$ (right), while the secondary image introduces additional wiggles.
            }
            \label{fig:time_bestfit}
        \end{minipage}\\[1.2em]
        \includegraphics[width=0.56\textwidth]{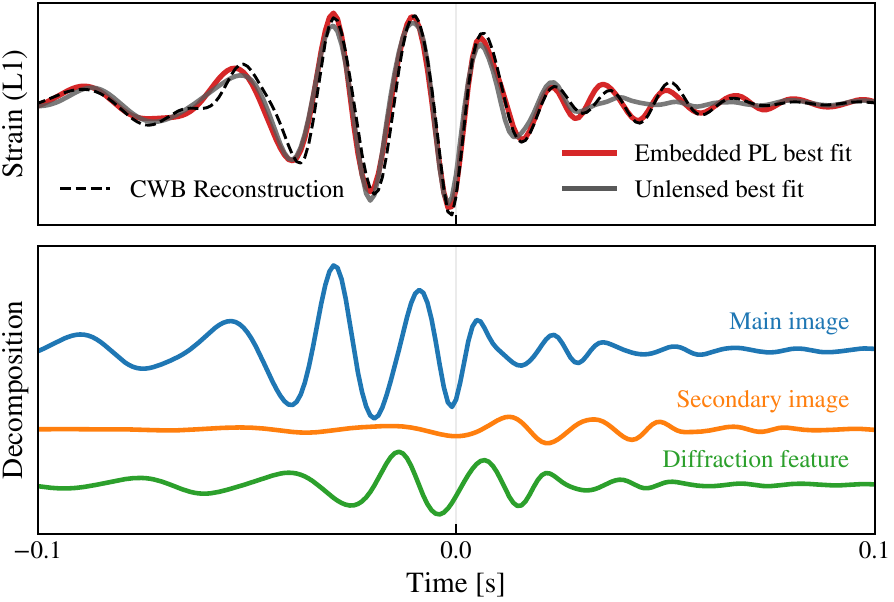}
        \caption{Time domain waveform for GW231123 (top) and decomposition of the best fit model into a main image, secondary image and diffraction feature (bottom). Lensed models refer to the best-fit source parameters and posterior mode where the time delay $\Delta t \sim 44$ms between the main and secondary microimages. This figure is an illustrative visualization of the best-fit waveform decomposition, and not used as a detection statistic.
        The embedded PL best fit agrees more closely with the model-independent reconstruction across detectors (L1 shown).}
        \label{fig:231123_lensed_time}
    \end{figure*}
    
\begin{figure*}
    \centering
    \includegraphics[width=0.85\linewidth]{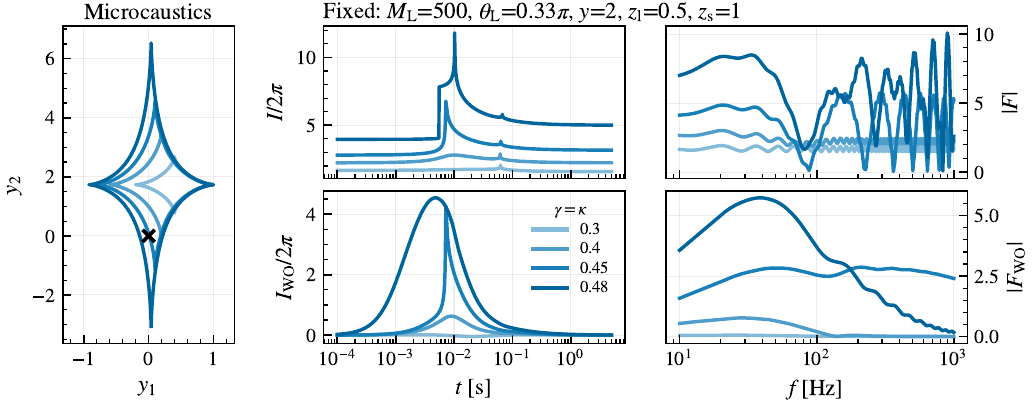}
    \includegraphics[width=0.85\linewidth]{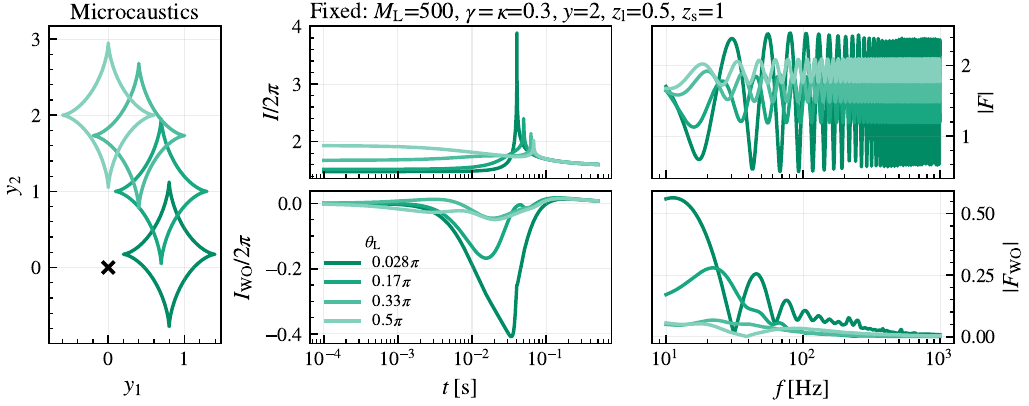}
    \caption{%
    Sensitivity of wave-optics diffraction to the external potential.
    Each panel shows the caustics associated to the microlens, the time-domain amplification $I(\tau)$ and the corresponding frequency-domain amplification factor $|F(f)|$, considering the total and diffractive contributions (labeled WO).
    Top: varying the external potential strength, assuming $\kappa=\gamma$, at fixed microlens position.
    Bottom: varying the microlens position angle at fixed external potential.
    Although macroscopic magnification is achromatic in geometric optics, the external potential modifies the microlensing caustics and time-delay response, producing frequency-dependent distortions in $F(f)$.
}
    \label{fig:pls_parameters_wo}
\end{figure*}

\section{Source properties, magnification and the mass gap}\label{app:source_properties}
The inferred source masses, spins, and distance under lensed and unlensed hypotheses, considering different source waveform models, in shown in table \ref{table:source}. Note that our unlensed analysis results are consistent with those of Ref. \citep{LIGOScientific:2025rsn}. For the embedded PL hypothesis, we assume $\kappa=\gamma$. We also show the sky localisation using Phenom waveforms, in Fig. \ref{fig:sky}. Under the lensing hypothesis, the sky localisation is broader and will be useful to search for the second image formed by strong lensing.

\begin{table*}[bh]
\centering
\renewcommand{\arraystretch}{1.5}
\begin{tabular}{lccc|ccc|ccc}
\hline
\hline
 &  & Unlensed  &  &  & Isolated PL & &  & Embedded PL &   \\
\hline
 & NRSur & Phenom & SEOBNR & NRSur & Phenom & SEOBNR & NRSur & Phenom & SEOBNR  \\
\hline
$\mathcal{M}^{\rm det}_c$ $[M_\odot]$ & $139^{+6.3}_{-8.4}$ & $118^{+9.7}_{-14}$ & $147^{+9.1}_{-9.1}$ & $123^{+14}_{-21}$ & $116^{+17}_{-26}$ & $133^{+15}_{-21}$ & $125^{+15}_{-20}$ & $118^{+18}_{-25}$ & $134^{+15}_{-18}$ \\
\hline
$q$ & $0.83^{+0.14}_{-0.16}$ & $0.6^{+0.13}_{-0.14}$ & $0.82^{+0.16}_{-0.21}$ & $0.53^{+0.28}_{-0.22}$ & $0.48^{+0.32}_{-0.2}$ & $0.61^{+0.25}_{-0.2}$ & $0.48^{+0.28}_{-0.17}$ & $0.48^{+0.36}_{-0.19}$ & $0.6^{+0.28}_{-0.22}$ \\
\hline
$D_L$ $[\rm Mpc]$ & $2099^{+1782}_{-998}$ & $930^{+392}_{-301}$ & $2446^{+1392}_{-984}$ & $5521^{+2085}_{-2173}$ & $4632^{+2456}_{-2093}$ & $5453^{+2120}_{-2472}$ & $8880^{+7234}_{-4181}$ & $7907^{+6537}_{-3884}$ & $7037^{+5831}_{-3292}$ \\
\hline
$\chi_{1}$ & $0.61^{+0.27}_{-0.38}$ & $0.73^{+0.19}_{-0.25}$ & $0.72^{+0.2}_{-0.33}$ & $0.49^{+0.35}_{-0.39}$ & $0.57^{+0.32}_{-0.46}$ & $0.51^{+0.34}_{-0.39}$ & $0.49^{+0.35}_{-0.35}$ & $0.56^{+0.34}_{-0.41}$ & $0.53^{+0.29}_{-0.4}$ \\
\hline
$\chi_{2}$ & $0.87^{+0.1}_{-0.28}$ & $0.42^{+0.42}_{-0.36}$ & $0.68^{+0.26}_{-0.45}$ & $0.34^{+0.49}_{-0.31}$ & $0.35^{+0.49}_{-0.32}$ & $0.41^{+0.46}_{-0.37}$ & $0.32^{+0.48}_{-0.28}$ & $0.32^{+0.5}_{-0.29}$ & $0.42^{+0.45}_{-0.38}$ \\
\hline
$a_1$ & $0.91^{+0.07}_{-0.25}$ & $0.79^{+0.17}_{-0.25}$ & $0.92^{+0.07}_{-0.21}$ & $0.77^{+0.2}_{-0.53}$ & $0.73^{+0.22}_{-0.48}$ & $0.81^{+0.16}_{-0.51}$ & $0.81^{+0.16}_{-0.39}$ & $0.7^{+0.25}_{-0.44}$ & $0.84^{+0.14}_{-0.5}$ \\
\hline
$a_2$ & $0.91^{+0.07}_{-0.26}$ & $0.66^{+0.3}_{-0.55}$ & $0.83^{+0.15}_{-0.45}$ & $0.46^{+0.47}_{-0.41}$ & $0.48^{+0.46}_{-0.43}$ & $0.55^{+0.4}_{-0.49}$ & $0.41^{+0.5}_{-0.36}$ & $0.45^{+0.47}_{-0.4}$ & $0.53^{+0.41}_{-0.47}$ \\
\hline
$\chi_p$ & $0.75^{+0.16}_{-0.19}$ & $0.73^{+0.19}_{-0.23}$ & $0.75^{+0.18}_{-0.22}$ & $0.5^{+0.34}_{-0.34}$ & $0.58^{+0.32}_{-0.4}$ & $0.53^{+0.32}_{-0.34}$ & $0.49^{+0.34}_{-0.32}$ & $0.57^{+0.33}_{-0.37}$ & $0.54^{+0.28}_{-0.34}$ \\
\hline
$\chi_{\rm eff}$ & $0.3^{+0.22}_{-0.25}$ & $0^{+0.18}_{-0.25}$ & $0.43^{+0.19}_{-0.25}$ & $0.36^{+0.22}_{-0.33}$ & $0.25^{+0.26}_{-0.3}$ & $0.39^{+0.23}_{-0.34}$ & $0.42^{+0.19}_{-0.31}$ & $0.24^{+0.27}_{-0.32}$ & $0.39^{+0.22}_{-0.33}$ \\
\hline
$m_{1}^{\mathrm{src}}$ $[M_\odot]$ & $126^{+17}_{-17}$ & $148^{+12}_{-12}$ & $131^{+18}_{-14}$ & $107^{+28}_{-22}$ & $113^{+27}_{-26}$ & $107^{+27}_{-21}$ & $93^{+29}_{-24}$ & $91^{+32}_{-25}$ & $98^{+29}_{-27}$ \\
\hline
$m_{2}^{\mathrm{src}}$ $[M_\odot]$ & $105^{+17}_{-18}$ & $89^{+19}_{-21}$ & $107^{+16}_{-22}$ & $56^{+22}_{-17}$ & $53^{+24}_{-19}$ & $65^{+27}_{-19}$ & $44^{+22}_{-14}$ & $44^{+24}_{-15}$ & $58^{+26}_{-17}$ \\
\hline
$\Msrctot$ $[M_\odot]$ & $232^{+25}_{-32}$ & $236^{+24}_{-26}$ & $237^{+26}_{-23}$ & $163^{+31}_{-23}$ & $166^{+32}_{-24}$ & $172^{+46}_{-28}$ & $137^{+40}_{-33}$ & $137^{+38}_{-34}$ & $157^{+42}_{-36}$ \\
\hline
$z$ & $0.38^{+0.25}_{-0.16}$ & $0.19^{+0.07}_{-0.06}$ & $0.43^{+0.2}_{-0.15}$ & $0.85^{+0.25}_{-0.28}$ & $0.73^{+0.3}_{-0.29}$ & $0.84^{+0.26}_{-0.33}$ & $1.2^{+0.78}_{-0.5}$ & $1.1^{+0.71}_{-0.48}$ & $1^{+0.65}_{-0.41}$ \\
\hline
$\log \mathcal{L}$ & $208^{+3.7}_{-5.9}$ & $205^{+3.9}_{-5.7}$ & $210^{+3.5}_{-5.4}$ & $211^{+3.9}_{-5.3}$ & $211^{+3.7}_{-5.3}$ & $211^{+3.8}_{-5.3}$ & $213^{+4.9}_{-5.7}$ & $212^{+4.5}_{-5.4}$ & $211^{+4.8}_{-4.8}$ \\

\hline
$\Delta \log \mathcal{L}_{\rm max}$ & $0$ & $-4.4$ & $0.5$ &$4.8$ & $1.9$ & $2.9$ & $6.3$ & $4.1$ & $5.8$  \\

\hline
\hline
\end{tabular}
\caption{Posterior medians of the source parameters of GW231123 binary black hole merger with 90\% credible intervals ($5-95$\%) for UL, Isolated PL, and Embedded PL hypotheses, using \texttt{NRSur7dq4} (NRSur), \texttt{IMRPhenomXPHM-ST} (Phenom)  and \texttt{SEOBNRv5PHM} (SEOBNR) waveform approximants. Lensing interpretation supports lower component masses and spins of the while increasing its redshift, consistently across the three waveform models. \label{table:source}}
\end{table*}
\begin{figure}[t]
    \centering
    \includegraphics[width=\linewidth]{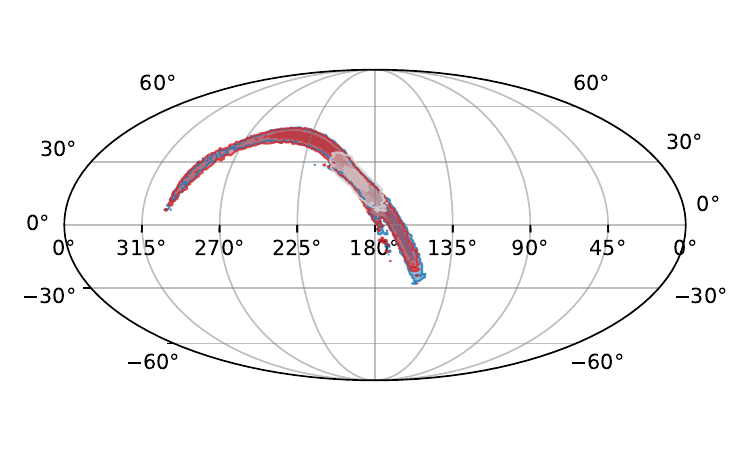}
    \caption{Sky localisation under Embedded PL (red), Isolated PL (blue) and unlensed hypothesis (gray) in the Phenom analysis.
    The lensing hypotheses yield broader localisations: relying on the more restricted unlensed sky map may prevent the identification of the host galaxy, additional macroimages, or electromagnetic counterparts in follow-up searches.}
    \label{fig:sky}
\end{figure}

%

Let us now explore macro-magnifications beyond the SIS relation and their impact on the astrophysical interpretation of the source and microlens. The pair-instability mass gap is a predicted absence of stellar-collapse black holes with masses between $\sim 60$ and $\sim 130\,M_\odot$: in this range, electron-positron pair production softens the core equation of state and triggers (pulsational) pair-instability supernovae, which partially or completely disrupt the star and prevent black-hole formation~\citep{Heger:2001cd,pisn} (the precise values depend on nuclear reaction rates \citep{Farmer:2020xne}).
Under the unlensed hypothesis, the components of GW231123 fall in or above the gap~\citep{LIGOScientific:2025rsn}, while the lensing hypotheses shift the inferred masses downwards (Table~\ref{table:source}).

The waveform-level inference constrains only the mass-sheet-invariant quantities ---the reduced shear $\tilde\gamma$ and the redshifted lens mass $M_{Lz}$--- while the absolute macro-magnification depends on the assumed macrolens through the convergence at the image position (Sec.~\ref{sec:lens_source_prop}).
To quantify this freedom, we apply a mass-sheet transformation to each posterior sample at fixed $\tilde\gamma$ and $M_{Lz}$: a convergence $\kappa$ rescales the apparent luminosity distance by $1/(1-\kappa)$, giving a total magnification $\mu = (1-\kappa)^{-2}/(1-\tilde\gamma^2)$; values below the pure-shear magnification ($\kappa=0$) correspond to negative convergence, i.e.~underdense lines of sight.
The true source redshift follows from the demagnified distance, lowering the source-frame masses as $\mu$ increases.

Figure~\ref{fig:magnification_vs_mass} shows the resulting source-frame component masses as a function of $\mu$.
The SIS closure ($\kappa=\gamma$) corresponds to a modest median magnification $\mu\simeq5$, for which the median primary mass, $92\,M_\odot$ (64--127 at 95\% c.l.), lies within the gap while the median secondary lies below it.
A magnification $\mu\simeq23$ ---about five times larger, attainable near the fold and cusp caustics of realistic elliptical lenses--- brings the median primary mass below the gap.
Larger magnifications also reduce the physical mass required of the microlens and boost the effective surface density of stellar-mass microlenses~\citep{Venumadhav:2017pps,Oguri:2017ock}, a mechanism observed for extremely magnified stars behind cluster caustics~\citep{Kelly:2017fps,Diego:2023qhp}.

\begin{figure}[t!]
    \centering
    \includegraphics[width=\linewidth]{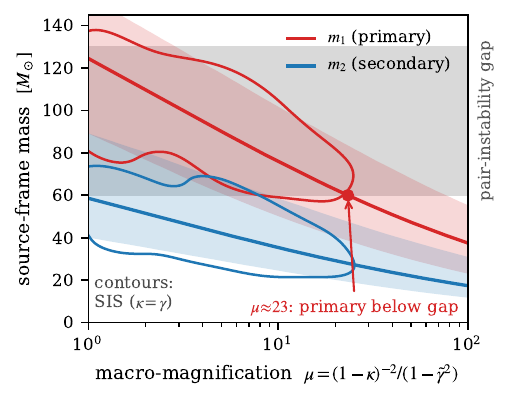}
    \caption{GW231123 source-frame component masses versus total macro-magnification $\mu$ (embedded PL, NRSur). Lines and shaded bands show the median and 95\% intervals as functions of $\mu$, obtained from mass-sheet transformations; the grey region marks the pair-instability mass gap. Contours show the 95\% c.l.\ two-dimensional posteriors of $(\mu, m_{1,2})$ under the $\kappa=\gamma$ (SIS) closure, which gives $\mu = 5.0^{+12.0}_{-3.9}$ (95\% c.l.). The dot marks $\mu\simeq23$, where the median primary mass falls below the gap.}
    \label{fig:magnification_vs_mass}
\end{figure}

The magnification therefore enters the astrophysical prior on the source: a population prior that penalizes the mass gap ---or, more generally, the high-mass tail of the black-hole distribution--- disfavors weakly magnified configurations and rewards strongly magnified ones, an effect that is absent when the magnification is fixed \emph{a priori}.
Astrophysical priors on GW231123 were recently examined in Ref.~\citep{Cheung:2026pky}, which combined a source prior from the observed population with a lens prior set by the lensing optical depth, for several \emph{isolated} lens models (point mass, generalized and cored isothermal spheres) and different physical origins for the microlens.
Their conclusions ---bounds on compact-object abundances still allow posterior odds favoring lensing, while predicted mass functions of black holes or collisionless cold dark matter halos disfavor it--- apply only to isolated lenses at fixed magnification.
For an embedded microlens the calculation changes qualitatively: the magnification becomes a free parameter of the macro-model (Fig.~\ref{fig:magnification_vs_mass}), the source-frame masses follow from it, and the relevant microlens abundance is that of a lens galaxy near a critical curve rather than a cosmological density of isolated objects~\citep{Venumadhav:2017pps,Diego:2019rzc}.
Extending the odds computation to embedded lenses, with priors on both the macro-model and the microlens population, is left for future work.

\section{Microlens mass and radius}\label{app:microlens_mass}

Although diffraction is sensitive to the redshifted microlens mass, $M_L(1+z_L)$, the intrinsic mass $M_L$ can be estimated with minimal additional assumptions.
To infer the intrinsic microlens mass we assign a lens redshift $z_L\!\in[0,z_S]$ to each posterior sample by weighting with the differential lensing probability. 
The weight combines the comoving volume element and the lensing cross section~\cite[Eqs.~68-70]{Tambalo:2022wlm},
\begin{equation}
    p(z_L\mid z_S) \propto 
    \frac{(1+z_L)^2}{H(z_L)}\,
    \frac{D_L D_{LS}}{D_S},
\end{equation}
which corresponds to a uniform mass function and a PL geometry. 

For each posterior sample we draw $z_L$ from this probability and convert the redshifted mass into $M$. 
Assuming an isothermal macrolens with $\kappa=\gamma$ and a flat lens mass function, the embedded PL NRSur analysis yields
$M_L\in(188,850)\,M_\odot$ (90\% c.l.), corresponding to $z_L\in(0.14,1.23)$.
The isolated PL case gives higher masses $M_L\in(468,1184)\,M_\odot$ and lower lens redshifts $z_L\in(0.09,0.66)$.

Our analysis is also valid for finite but compact lenses. We can estimate the maximum radius such that the results converge to a PL by requiring the formation of at least two microimages. For $y\gtrsim 1$ (as for the embedded PL), the second microimage is located at $x_-\sim 1/y$.
For a homogeneous lens with physical radius $R_L$, the existence of this image requires that the integrated mass at the position of this image is $\sim M_L$~\citep{Zumalacarregui:2024ocb}. Introducing the physical scale from the posterior samples implies an upper limit $R_L<\xi_0/y$, where $\xi_0$ is the Einstein radius in the lens plane. 
Using the posterior distribution of $y$ (rescaled for the SIS macro-model) we find 
$R_L<0.35\,\mathrm{pc}$ at 95\% confidence for an isolated object.
Analyses including extended or embedded lens profiles will provide more robust constraints on the mass profile and physical size of the lens, as recently explored for isolated lenses~\citep{Cheung:2026pky}.

\section{Microlens as dark matter}
\label{app:microlens_abundance}


We constrain the abundance of compact microlenses by treating the detection of GW231123 as a Poisson process with one observed event~\citep{Zumalacarregui:2017qqd,Basak:2021ten,Gais:2022xir,barsode2024constraintscompactdarkmatter}. 
The probability of observing $k$ lensed events given a fractional abundance $f_{\rm C}$ and total exposure $\lambda_{\rm tot}$ is
\begin{equation}
    p(k \mid f_{\rm C},\lambda_{\rm tot}) = \frac{(f_{\rm C}\lambda_{\rm tot})^k}{k!}\,e^{-f_{\rm C}\lambda_{\rm tot}}\,.
\end{equation}
Here $f_{\rm C} \equiv \Omega_{\rm C}/\Omega_{\rm cdm}$ 
is the fraction of dark matter in compact objects, assumed constant with redshift and homogeneously distributed.
We take $k=1$, corresponding to GW231123, and compute the posterior for $f_{\rm C}$ using a flat prior.

The total exposure $\lambda_{\rm tot}$ is the expected number of detected microlensing events if $f_{\rm C}=1$,
\begin{equation}
    \lambda_{\rm tot} = \tilde\lambda_{\rm L} + \tilde\lambda_{\rm UL}\,,
\end{equation}
where $\tilde\lambda_{\rm L}$ is the contribution from the detected event (L) and $\tilde\lambda_{\rm UL}$ that of all remaining unlensed (UL) events.  
We distinguish $\lambda=f_{\rm C}\tilde\lambda$ to factor out the dependence on $f_{\rm C}$ explicitly.

For a source at redshift $z_s$, the expected number of detectable microlenses is
\begin{equation}
    \tilde\lambda(z_s) = \int_0^{z_s}\! dz_l\, n(z_l)\,\sigma(z_l)\,,
\end{equation}
where $n(z_l)=\rho_{\rm cdm}(z_l)/M_L$ is the number density of compact objects (for $f_{\rm C}=1$) and $\sigma(z_l)=\pi[\xi_0(z_l) y_{\rm cr}]^2$ is the microlensing cross section in the lens plane. 
This quantity is closely related to the optical depth $\tau(z_s)$, but differs by factors of $(1+z_l)$ and $H(z_l)$ that account for the differential lensing volume per source and are already absorbed into the definition of the convergence $\kappa(z_s)$.
The convergence represents the projected surface density of compact objects in units of the critical surface density, such that
\begin{equation}
    \tilde\lambda = \kappa\,y_{\rm cr}^2\,.
\end{equation}
In this form, $\kappa$ gives the mean number of microlenses within one Einstein radius of the line of sight, while $y_{\rm cr}$ defines the maximum detectable impact parameter (in units of the Einstein radius).  This identity can be understood by noting that $\kappa \sim n\,\pi\xi_0^2 D_l$ measures the surface filling factor of Einstein disks, so that multiplying by $y_{\rm cr}^2$ gives the expected number of microlenses within a disk of radius $y_{\rm cr}\xi_0$ around a given source.

For GW231123 we compute $\tilde\lambda_{\rm L}=\kappa_{\rm L}y_{\rm L}^2$ directly from the posterior samples of the embedded PL analysis, where $\kappa_{\rm L}$ is the convergence of the SIS macro-model and $y_{\rm L}$ is the corresponding (rescaled) source–microlens offset. 
For unlensed events we approximate the exposure of each source as $\tilde\lambda_i=\bar\kappa(z_i)y_{\rm cr}^2$, where $\bar\kappa(z)$ is the average convergence to redshift $z$ (Eq.~(A9) of Ref.~\citep{Zumalacarregui:2017qqd}). 
Following the detectability studies of Ref.~\citep{mishrapopbiases2023}, we adopt $y_{\rm cr}=1.5$ for these events and sum over all confident GWTC-4 detections ($p_{\rm astro}>0.5$, ${\rm FAR}<1\,{\rm yr}^{-1}$)~\citep{LIGOScientific:2025slb}, obtaining
\begin{equation}
    \tilde\lambda_{\rm UL} \simeq 7.0\,\left(\frac{y_{\rm cr}}{2}\right)^2\,.
\end{equation}

For each posterior sample of GW231123, we compute $\lambda_{\rm tot}=\tilde\lambda_{\rm L}+\tilde\lambda_{\rm UL}$ and evaluate
\begin{equation}
    p(f_{\rm C} \mid k=1,\lambda_{\rm tot}) \propto (f_{\rm C}\lambda_{\rm tot})\,e^{-f_{\rm C}\lambda_{\rm tot}},\qquad 0\le f_{\rm C}\le1\,,
\end{equation}
which is a truncated Gamma distribution with shape parameter $k+1=2$ and rate $\lambda_{\rm tot}$. 
Sampling $f_{\rm C}$ from this posterior for each realization yields a joint distribution of $(M,f_{\rm C})$ values, shown in Fig.~\ref{fig:dark_matter_after_GW231123}, from which the confidence intervals on the microlens mass and abundance are derived.
The figure also shows excluded regions (95\% c.l.) by lensed quasars~\citep{Esteban-Gutierrez:2023qcz}, stars~\citep{Mroz:2024mse,Blaineau:2022nhy,Oguri:2017ock,Diego:2017drh}, type Ia supernovae \citep{Zumalacarregui:2017qqd,DES:2024ffp}, as well as dynamics of ultra-faint dwarf galaxies~\citep{Brandt:2016aco,Graham:2024hah}, assuming point-like objects. 
The cosmic microwave background sets additional limits on compact objects formed before recombination~\citep{Serpico:2020ehh,Agius:2024ecw,Croon:2024rmw}. 

\begin{figure}[t!]
    \centering
    \includegraphics[width=0.99\linewidth]{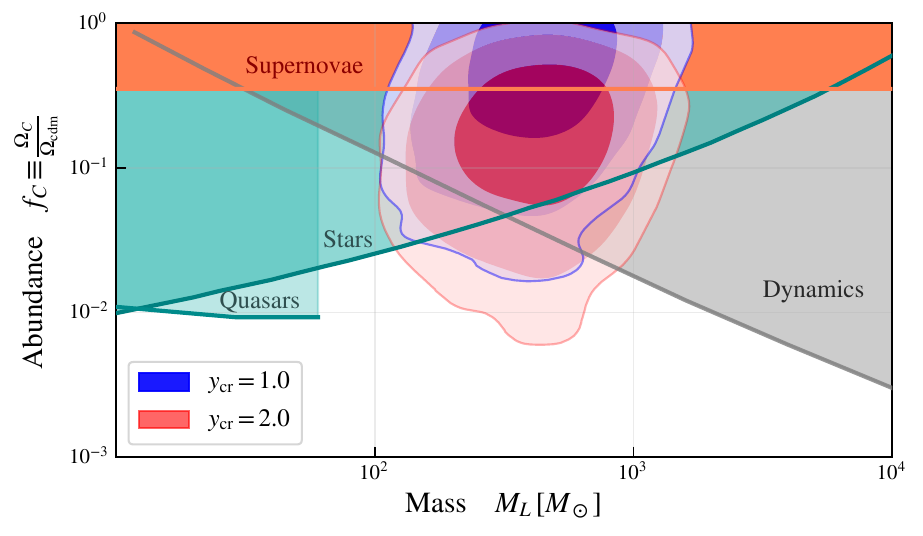}
    \caption{Dark matter after GW231123. Contours show the mass and abundance of microlenses derived from the Embedded PL analysis (NRSur, 68, 95 \& 99\% confidence regions). Shaded areas show the regions excluded by various probes (95\% confidence). The inferred microlens abundance is in 1--2$\sigma$ tension with dynamical limits and stellar microlensing (see text).} 
    \label{fig:dark_matter_after_GW231123}
\end{figure}

The interpretation of the microlens as a dark-matter object is in tension with existing limits at the level of $\sim 1\, (2)\sigma$ for $y_{\rm cr} = 2\, (1)$ in the mass range $M_L\sim 200-1,000 M_\odot$. The tension may be reduced in the case of extended dark-matter objects, for which the limits become less stringent~\citep{Croon:2024jhd}.
Note that stellar lensing probes the Milky-Way halo and dynamical bounds ultra-faint dwarf galaxies: the corresponding limits would be lifted if dark-matter object formation has environmental dependence, e.g. velocity-dependent self-interactions~\citep{Gilman:2021sdr}.
A related possibility is that the microlens is a dark-matter halo compact enough to mimic a point lens, rather than a point-like object. Ref.~\citep{Cheung:2026pky} finds that collisionless cold dark matter does not form halos concentrated enough to account for GW231123, whereas self-interacting dark matter with a large cross section at low velocities can, through gravothermal collapse; the resulting posterior odds are inconclusive for that scenario. These estimates assume an isolated lens, and would need to be revised in the presence of an external potential, which lowers both the required halo mass and the abundance needed to account for the event, the latter as $f_{\rm c}\propto(1-\kappa)\tilde y^{-2}$ (see below).
Finally, it is important to remember that the required abundance depends on the physical impact parameter as $f_{\rm c}\propto y^{-2} \propto (1-\kappa)\tilde y^{-2}$ due to the mass sheet degeneracy (Eq.~\ref{eq:F_mass_sheet}). Our analysis assumed $\kappa=\gamma$ (Eq.~\ref{eq:macro_SIS}), but the abundance can be made lower if the macroimage forms in a region where $\kappa \sim 1$. 
Future GW observations will vastly improve our capacity to probe dark-matter objects thanks to the richness of wave-optics lensing phenomena.


\section{Alternatives to a single heavy microlens}
\label{app:microlens_alternatives}

The waveform-level analysis constrains the effective lensing mass scale $M_{Lz}$ and locally-produced wave-optics features, but does not establish that the microlens is a single compact object.
Here we illustrate two astrophysically motivated alternatives that closely reproduce the wave-optics signature of a heavy point lens: a single extended lens, such as the central region of a stellar cluster~\citep{Savastano:2023spl,Cheung:2024ugg}, and the collective effect of many stellar-mass objects~\citep{Diego:2019rzc,Shan:2024min,Palencia:2023kne,Zumalacarregui:2026uqs}.

\begin{figure*}[p]
    \centering
    \includegraphics[width=\linewidth]{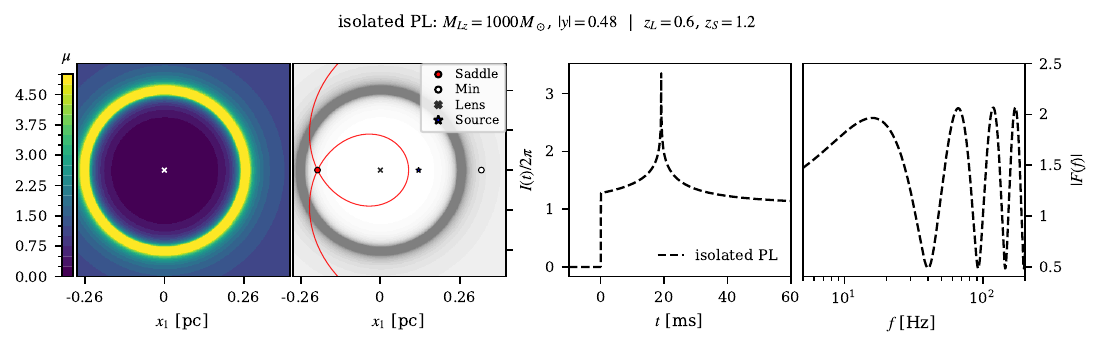}\\[0.4em]
    \includegraphics[width=\linewidth]{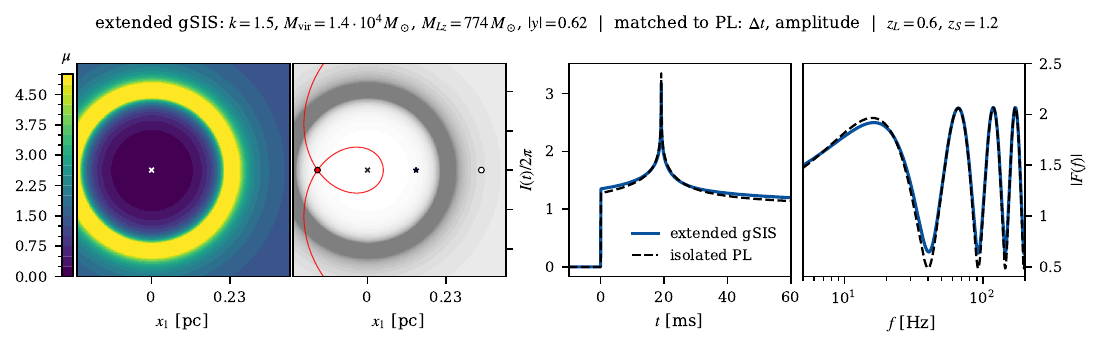}\\[0.4em]
    \includegraphics[width=\linewidth]{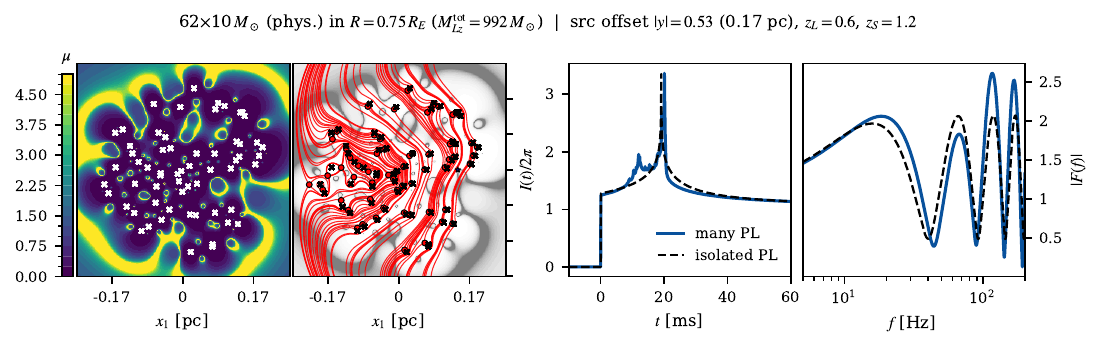}
    \caption{Alternatives to a single heavy microlens (isolated configurations at $z_L=0.6$, $z_S=1.2$), in order of increasing complexity.
    Each row shows the magnification map and time-delay contours in the lens plane (left panels, physical units; microlens positions, image positions and source as labeled), the time-domain lensing response $I(t)$, and the amplification factor $|F(f)|$ (right panels); the isolated point lens (black dashed) is repeated in all rows for comparison.
    \emph{Top}: fiducial isolated point lens with $M_{Lz}=10^3\,M_\odot$ and image delay $\Delta t \approx 20$\,ms, comparable to the GW231123 wave-optics feature.
    \emph{Middle}: extended gSIS lens ($k=3/2$) matched to the point lens' delay and peak amplification: effective mass $M_{Lz}\simeq 770\,M_\odot$, but total (virial) mass $M_{\rm vir}\simeq 1.4\times10^4\,M_\odot$.
    \emph{Bottom}: 62 point lenses of $10\,M_\odot$ within $0.75\,R_E$ (same total mass and source offset): the stellar field reproduces the heavy lens' response, resolving the saddle spike into micro-caustic features.
    }
    \label{fig:microlens_alternatives}
\end{figure*}

Figure~\ref{fig:microlens_alternatives} compares three isolated configurations of increasing complexity:
a fiducial point lens with $M_{Lz}=10^3\,M_\odot$ and impact parameter $y=0.48$, chosen to produce an image delay $\Delta t \approx 20$\,ms, comparable to the wave-optics feature inferred for GW231123 (Fig.~\ref{fig:time_corner}); an extended cluster; and a collection of light point lenses.
For illustration, the lenses are placed at $z_L=0.6$ and the source at $z_S=1.2$; an external potential is omitted for simplicity.
The first row considers an isolated PL, showing the image-plane magnification, the isochrones associated to type-II microimages, the time-domain lensing response $I(\tau)$ and frequency-domain amplification factor (absolute value) $|F(f)|$.

The second row of Fig.~\ref{fig:microlens_alternatives} shows an extended lens: a generalized singular isothermal sphere (gSIS) with density profile $\rho\propto r^{-(k+1)}$ and slope $k=3/2$, describing, e.g., the dense inner region of a stellar cluster~\cite[Sec.~IIIA]{Tambalo:2022wlm}.
Matching the point lens' time delay and peak amplification requires an effective mass $M_{Lz}\simeq 770\,M_\odot$ at a slightly larger offset ($y=0.62$); the resulting amplification factor is then nearly indistinguishable from the point lens over the sensitive band.
The effective mass probed by diffraction is a small fraction of the total (virial) mass of the extended lens, see Eq.~(32) of \citet{Tambalo:2022wlm}.
For the configuration shown ($k=3/2$, $M_{Lz}\simeq 770\,M_\odot$) this gives $M_{\rm vir}\simeq 19\,M_{Lz}\simeq 1.4\times10^4\,M_\odot$, which is on the light end of globular clusters~\citep{Baumgardt:2018pyl}.
The total mass grows steeply for shallower profiles: an isothermal sphere ($k=1$) with the same effective mass has $M_{\rm vir}\simeq 2.2\times10^3\,M_{Lz}\simeq 1.7\times10^6\,M_\odot$, comparable to the high-mass end of a globular cluster.

The third row shows 62 objects of $10\,M_\odot$ each (matching the total redshifted mass, $\simeq 990\,M_\odot$), distributed uniformly within $0.75\,\xi_0$ ($\simeq 0.2$\,pc) of the heavy lens' Einstein radius, with the same physical source offset.
The time-domain response closely tracks that of the single heavy lens, with the saddle-point singularity resolved into a cluster of micro-caustic features spanning $\sim 8$--$21$\,ms.
Across the observable band the amplification factor exhibits the same oscillatory structure and amplitude. 
A sufficiently dense stellar field can therefore mimic a single heavy microlens; in strongly magnified macroimages the boosted effective stellar surface density makes such configurations more likely~\citep{Venumadhav:2017pps,Diego:2019rzc,Palencia:2023kne}.
A systematic exploration at the level of parameter estimation requires an effective description of stellar-field diffraction~\citep{Zumalacarregui:2026uqs} and will be presented elsewhere.

These examples show that the diffraction signatures inferred for GW231123 may be compatible with astrophysically ordinary configurations---dense stellar fields or star clusters---in addition to a single intermediate-mass compact object.
The three configurations are nearly degenerate at current sensitivity.
Discriminating among them requires modeling secondary diffraction features (e.g.~the micro-caustic structure in the third row of Fig.~\ref{fig:microlens_alternatives}), the central images of broad profiles ($k<1$), or additional lensing signatures, motivating the development of stochastic and extended-lens wave-optics analyses~\citep{Zumalacarregui:2026uqs,Cheung:2024ugg,Feldbrugge:2019fjs,Jow:2022pux,Feldbrugge:2026qux}.


\bibliography{gw_lensing}

\end{document}